\documentclass{ieeeaccess}
\usepackage{cite}
\usepackage{amsmath,amssymb,amsfonts}
\usepackage{graphicx}
\usepackage{textcomp}
\usepackage{subcaption}
\usepackage[capitalize]{cleveref}
\crefname{appsec}{Appendix Section}{Appendix Sections}  
\crefname{appsubsec}{Appendix Subsection}{Appendix Subsections}  

\usepackage{xcolor}
\usepackage{multicol}
\usepackage{cuted} 
\usepackage{algorithm}
\usepackage{algpseudocode}

\usepackage{graphicx}
\usepackage{subcaption} %
\usepackage{multirow}
\usepackage{booktabs}
\usepackage{enumitem}

\usepackage[acronyms,nonumberlist,nopostdot,nomain,nogroupskip,acronymlists={hidden}]{glossaries} 
\newglossary[algh]{hidden}{acrh}{acnh}{Hidden Acronyms}

\def\BibTeX{{\rm B\kern-.05em{\sc i\kern-.025em b}\kern-.08em
    T\kern-.1667em\lower.7ex\hbox{E}\kern-.125emX}}

\DeclareMathOperator*{\argmin}{arg\,min}

\DeclareMathOperator*{\argmax}{arg\,max}

\begin{document}

\newacronym{fmea}{FMEA}{Fault Mode and Effect Analysis}
\newacronym{dc}{DC}{Diagnostic Coverage}
\newacronym{rv}{RV}{Random Variable}
\newacronym{pdf}{PDF}{Probability Density Function}
\newacronym{pt}{PT}{Partial Test}
\newacronym{ttf}{TTF}{Time-to-Failure}
\newacronym{hpp}{HPP}{Homogenous Poisson Process}
\newacronym{nhpp}{NHPP}{Non-Homogenous Poisson Process}
\newacronym{kld}{KLD}{Kullback-Leibler Divergence}
\newacronym{shd}{SHD}{Squared Hellinger Distance}
\newacronym{auc}{AUC}{Area Under the Curve}
\newacronym{ateer}{ATEER}{Absolute Total Expected Event Error}
\newacronym{mse}{MSE}{Mean Squared Error}
\newacronym{ccf}{CCF}{Common Cause Failures}
\newacronym{mle}{MLE}{Maximum Likelihood Estimation}
\newacronym{fmea}{FMEA}{Failure Mode and Effects Analysis}
\newacronym[longplural={Fisher Information Matrices}]{fim}{FIM}{Fisher Information Matrix}
\newacronym{sdp}{SDP}{Semidefinite Program}
\newacronym{misdp}{MISDP}{Mixed Integer Semidefinite Program}
\newacronym{al}{AL}{Active Learning}
\newacronym{mc}{MC}{Monte Carlo}
\newacronym{nasa}{NASA}{National Aeronautics and Space Administration}
\newacronym{it}{IT}{Information Technology}
\newacronym{dl}{DL}{Deep Learning}
\newacronym{ml}{ML}{Machine Learning}

\newacronym{egra}{EGRA}{Efficient Global Reliability Analysis}
\newacronym{akmcs}{AK-MCS}{Active Kriging Monte Carlo Simulation }
\newacronym{sis}{SIS}{Safety instrumented systems}
\newacronym{pfd}{PFD}{probability of failure on demand}
\newacronym{alc}{ALC}{Active Learning Curves}
\newacronym{bald}{BALD}{Bayesian Active Learning by Disagreement}
\newacronym{mttf}{MTTF}{Mean-Time-to-Failure}
\newacronym{af}{AF}{Acquisition Function}

\history{Date of publication xxxx 00, 0000, date of current version xxxx 00, 0000.}
\doi{10.1109/ACCESS.2017.DOI}

\title{\normalsize Active Learning for Repairable Hardware Systems with Partial Coverage}
\author{\uppercase{Michael L. Potter\authorrefmark{1}, \IEEEmembership{Student, IEEE},
Beyza Kalkanli \authorrefmark{1}, 
\IEEEmembership{Student, IEEE}, 
\uppercase{Deniz Erdogmus \authorrefmark{1}, 
\IEEEmembership{Member, IEEE}, and Michael Everett}\authorrefmark{1},
\IEEEmembership{Member, IEEE}}
\address[1]{Northeastern University, Boston, MA 02115 USA}
\tfootnote{© 2025 IEEE. Personal use of this material is permitted. Permission from IEEE must be obtained for all other uses, in any current or future media, including reprinting/republishing this material for advertising orpromotional purposes, creating new collective works, for resale or redistribution to servers or lists, or reuse of any copyrighted component of this work in other works. Submitting to IEEE Access Journal - Reliability Society}
}
\markboth
{Author \headeretal: Preparation of Papers for IEEE TRANSACTIONS and JOURNALS}
{Author \headeretal: Preparation of Papers for IEEE TRANSACTIONS and JOURNALS}

\corresp{Corresponding author: Michael Potter (email: potter.mi@northeastern.edu)}

\begin{abstract}
Identifying the optimal diagnostic test and hardware system instance to infer reliability characteristics using field data is challenging, especially when constrained by fixed budgets and minimal maintenance cycles.  \gls{al} has shown promise for parameter inference with limited data and budget constraints in machine learning/deep learning tasks. However, \gls{al} for reliability model parameter inference remains underexplored for repairable hardware systems. It requires specialized \gls{al} \glspl{af} that consider hardware aging and the fact that a hardware system consists of multiple subsystems, which may undergo only partial testing during a given diagnostic test. 
To address these challenges, we propose a relaxed \gls{misdp} \gls{al} \gls{af} that incorporates \gls{dc}, \glspl{fim}, and diagnostic testing budgets. Furthermore, we design empirical-based simulation experiments focusing on two diagnostic testing scenarios: (1) partial tests of a hardware system with overlapping subsystem coverage, and (2) partial tests where one diagnostic test fully subsumes the subsystem coverage of another. We evaluate our proposed approach against the most widely used \gls{al} \gls{af} in the literature (entropy), as well as several intuitive \gls{al} \glspl{af} tailored for reliability model parameter inference. Our proposed \gls{af} ranked best on average among the alternative \glspl{af} across 6,000 experimental configurations, with respect to \gls{auc} of the \gls{ateer} and \gls{mse} curves, with statistical significance.
\end{abstract}


\glsresetall

\titlepgskip=-15pt

\maketitle

\section{Introduction}
Quantifying the intrinsic reliability of field-deployed hardware systems is a critical research field, with applications in aerospace \cite{lindsey2020reliability}, computer architectures \cite{de2024microsoft}, automotive industry \cite{davis2003reliability}, medical devices \cite{abd2023critical}, and military equipment \cite{potterweapon}. For example, aerospace organizations face significant risks to both financial gains and human life due to rocket malfunctions \cite{latimes2015spacestation}. In the automotive industry, companies have experienced injuries and fatalities from faulty systems such as defective throttles, airbags, and ignition switches \cite{tuskr2024automotivetesting}. In computing, a major global \gls{it} disruption occurred in 2024 to a faulty configuration update \cite{messageware2024crowdstrikeoutage}. Medical devices, including pacemakers and defibrillators, have been recalled for circuit and structural failures, impacting both patient safety and financial stability \cite{medicaldevice2024recalls}. Lastly, in military applications, the functionality of equipment such as tanks, convoys, and missiles is crucial for mission success and national security, particularly with prolonged use or maintenance gaps \cite{potter2024bayesian}.

To combat these risks, reliability analysis is a standard component in product development, maintenance strategies, quality assurance and risk analysis \cite{relautomotiveaero,medicaldevice2024qa}. However, with limited time and budget to gather field data on hardware systems in operational environments, prioritizing which data to collect is crucial for accurately assessing the intrinsic reliability of a system. Therefore, reliability analysis faces several key challenges: developing cost-effective diagnostic tests that detect a wide range of failure modes and events, accurately modeling these failure modes/events when specialized tests are used, and selecting the most appropriate diagnostic test to infer the reliability characteristics using the collected data as quickly and cost-effectively as possible. This work focuses on addressing the second and third challenges. 

The optimal data selection challenge has been extensively studied in the \gls{al} field, particularly for classification and regression tasks.
Given a large pool of unlabeled data and a labeling oracle, \gls{al} algorithms aim to infer model parameters that maximize accuracy using the smallest labeled dataset \cite{settles2009active,moustapha2022active}. In reliability analysis, \gls{al} has been primarily applied to structural reliability, which slightly diverges from the typical \gls{al} objective \cite{li2021active}. Structural reliability focuses on constructing cost-effective surrogates for the limit-state function, aiming to accurately approximate the boundary between acceptable and unacceptable structural performance while minimizing model evaluations \cite{BHATTACHARYA2022169,moustapha2022active}. Notable surrogate models include Gaussian processes, with \gls{egra} \cite{bichon2008efficient} and \gls{akmcs} \cite{echard2011ak} as key examples. We apply \gls{al} for reliability model inference; however, unlike \gls{al} for structural reliability, our goal is not to approximate the boundary between acceptable and unacceptable structural performance in a covariate space using surrogate models with the minimal surrogate model evaluations. Instead, we aim to learn the reliability model parameters by collecting real-world field data from the hardware system during operational use.

Assigning failure events to the correct subsystems enhances reliability likelihood formulation by enabling more granular block diagram representations of hardware systems. When prior knowledge about subsystem failure intensities is limited, hardware systems are typically modeled as combinations of independent parallel and series subsystems \cite{vcepin2011assessment}. These block diagrams facilitate reliability allocation by identifying and improving low-reliability subsystems to achieve overall system reliability \cite{kuo2007recent}. With additional domain knowledge, such as diagnostic tests' \gls{dc} (where \gls{dc} measures the failure detection rate and isolation of a diagnostic test), subsystem failure rates can be derived and related through \gls{dc} coefficients, contributing to the total failure rate. In \gls{sis}, this approach is used to approximate the overall \gls{pfd} as the sum of \gls{pfd} from functional and diagnostic testing, linked via the system’s total failure rate or viewed as serial subsystem reliability \cite{lundteigen2007effect}.  However, most works on \gls{sis} estimate \gls{dc} by determining a time independent failure intensity (typically \gls{mttf}) through empirical methods or \gls{mle}, often relying on look-up tables \cite{maranzanomaximum,maranzanosubystem}.
Furthermore, it has been shown that combining targeted subsystem and hardware system diagnostic tests can achieve the same \gls{mse} on reliability parameter estimates as whole-hardware system testing, but with lower costs (for a given modeling mismatch quantification).  

While \cite{maranzanomaximum,maranzanosubystem} developed robust subsystem test plans that balance estimation accuracy, modeling error, and cost, they do not address time-varying systems, sequential diagnostic test selection, or \gls{dc}. Our approach addresses these gaps by designing a more accurate reliability likelihood that incorporates \gls{dc} during parameter inference for time-dependent models. It also optimizes diagnostic test schedules to efficiently select subsystems for testing, balancing budget constraints and field data availability.

Overall, this paper introduces an improved \gls{al} \gls{af} that incorporates \gls{dc} to improve reliability model parameter inference using minimal field data from time-dependent repairable hardware systems. By integrating \gls{dc} into the reliability model likelihood specification, our method leverages diagnostic test information to enhance parameter estimation. We formulate a relaxed \gls{misdp} that incorporates the \gls{fim} of the model reliability likelihood under diagnostic tests which only partially test the hardware system and budget constraints. To the best of our knowledge, the proposed approach is the first \gls{af} for reliability modeling to integrate continuous-time reliability models, \gls{dc}, and budget constraints. Our experiments demonstrate that widely adopted \gls{al} strategies, such as entropy that typically excels in other domains, struggle to effectively select diagnostic tests to perform at a maintenance cycle. In contrast, our proposed optimization strategy for maintenance tests, outlined in \cref{subsec:proposedstrategy}, significantly outperforms other \gls{al} methods with statistical significance.




\label{sec:introduction}
\begin{figure}[h]
    \centering
    \includegraphics[width=0.8\linewidth]{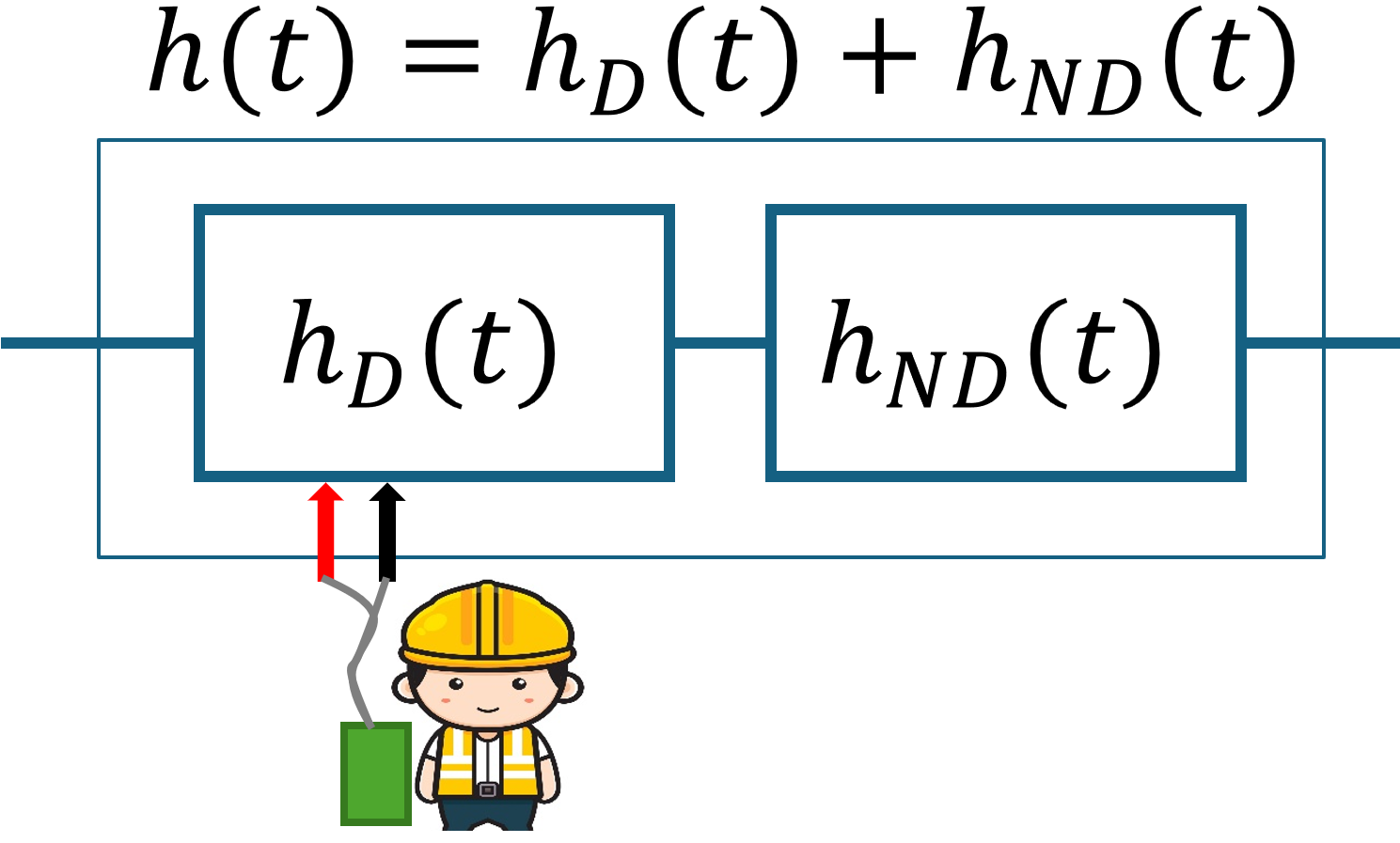}
    \caption{Repairman Bob performs a diagnostic test which only tests a subsystem of a hardware system}
    \label{fig:oraclebob}
\end{figure}

\section{Problem Statement}
Our objective is to minimize the number of maintenance cycles required to infer reliability model parameters while incorporating the following information: 
\begin{enumerate}
    \item a diagnostic testing budget at each maintenance cycle
    \item the \gls{dc} of diagnostic tests
    \item the aging and repairable characterisitcs of a hardware system
\end{enumerate}

Theoretically, reliability model inference using data from subsystem-level diagnostic tests can match the \gls{mse} of full testing on reliability model parameters, but at a lower data collection cost, for a desired \gls{mse} and a specified modeling mismatch error \cite{maranzanomaximum,mcshane2008count}. This makes subsystem testing a cost-efficient alternative. However, selecting the optimal diagnostic test for each maintenance cycle is critical to realizing this efficiency. Drawing inspiration from \cite{maranzanosubystem,sourati2017probabilistic,zhang2000value}, we select diagnostic tests by maximizing the trace of the \gls{fim} of the reliability likelihood function, under budget constraints and incorporating the knowledge of \gls{dc}.

\section{Related work}
Active Learning (\gls{al}) has gained attention for its potential to reduce data collection costs by selecting the most informative samples \cite{settles2009active}. In reliability engineering, \gls{al} has primarily been applied to structural reliability problems, focusing on static systems where the failure surface is approximated using surrogate models such as Gaussian Processes. Methods like Active Kriging Monte Carlo Simulation (\gls{akmcs}) and Efficient Global Reliability Analysis (\gls{egra}) are widely used for this purpose \cite{echard2011ak,bichon2008efficient}. These approaches significantly reduce computational costs but are ill-suited for systems with dynamically changing failure patterns, as they do not incorporate time-dependent reliability metrics.  

In machine learning, \gls{al} techniques are often driven by uncertainty-based sampling strategies, such as entropy and margin sampling \cite{settles2009active}, mutual information \cite{sourati2016mutual}, and \gls{bald} \cite{houlsby2011bald}. Extensions to regression tasks, like Bayesian optimization, address challenges such as non-uniform data distributions and sparse sampling regions \cite{snoek2012practical}. However, these methods do not incorporate domain-specific priors, such as \gls{dc}, nor do they account for the interdependencies between test selection and reliability model inference, which limits their efficiency and applicability in reliability analysis.

Recent research has sought to bridge this gap by tailoring \gls{al} for reliability engineering. For instance, a new active learning method for system reliability analysis with multiple failure modes combines subset simulation and Kriging-based surrogate modeling to dynamically update failure regions and improve accuracy \cite{moran2023variance}. Similarly, a reliability analysis approach for arbitrary systems integrates active learning and global sensitivity analysis to handle complex systems with multiple failure modes \cite{moustapha2024arbitrary}. In a related domain, the work on supervised learning for coverage-directed test selection in simulation-based verification highlights the importance of incorporating \gls{dc} into active learning pipelines \cite{masamba2022supervised}. This principle has inspired reliability engineering applications, such as integrating \gls{dc} into active learning pipelines to optimize data acquisition. These advancements have broadened the applicability of \gls{al} to dynamic and complex systems.  However, unlike these methods, which typically focus on surrogate models and sensitivity analysis, our approach explicitly integrates domain-specific priors, such as \gls{dc}, and accounts for the interdependencies between test selection and reliability model inference, offering a more comprehensive solution for reliability analysis.

\begin{figure}
    \centering
    \includegraphics[width=0.9\linewidth]{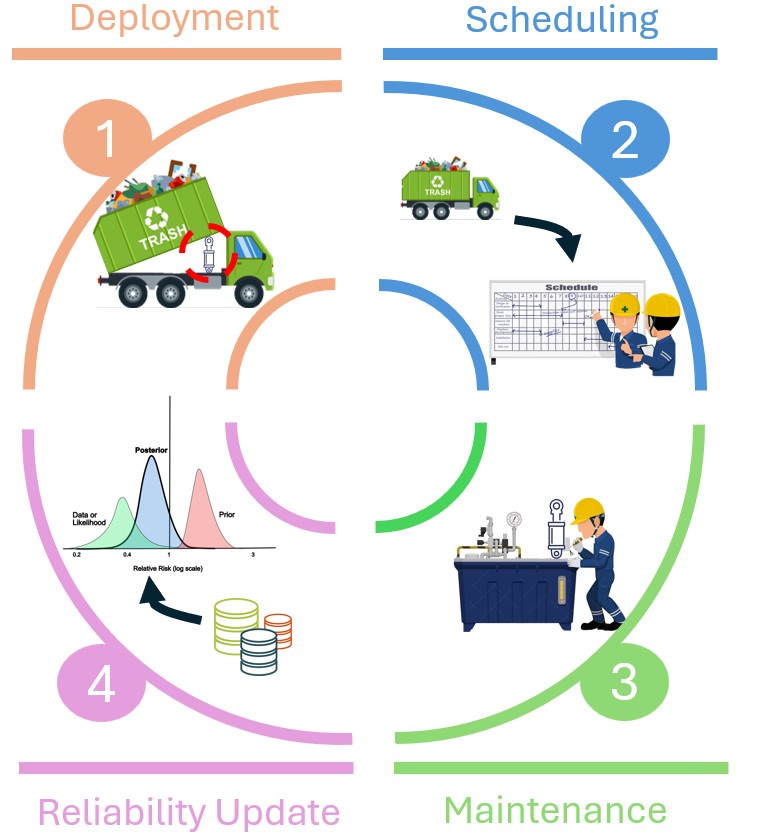}
    \caption{The cycle of reliability analysis for a repairable hardware system.}
    \label{fig:cycleofreliability}
\end{figure}

\section{Preliminaries}
The maintenance life cycle of repairable hardware systems consists of four stages: (1) deployment of the system into the field, (2) scheduling of preventive maintenance or inspections, (3) performing maintenance and inspections, and (4) updating reliability models using newly acquired data (\cref{fig:cycleofreliability}) \cite{assetlifecycle}. Understanding this maintenance life cycle is essential for accurately modeling failure event data and optimizing maintenance strategies.

In the following subsections, we detail our assumptions for modeling failure event data in repairable hardware systems. These assumptions include \gls{dc} with time-varying failure intensities, These assumptions include \gls{dc} with time-varying failure intensities, diagnostic tests where two tests may partially overlap or one may completely cover the other (overlapping or subset diagnostic tests respectively), and the dataset formulation, and dataset formulation. To simplify terminology, we use the terms maintenance, maintenance test, and repair interchangeably throughout the paper.

\subsection{Failure Modeling}
\label{subsec:failure_modeling}
We model a hardware system as a repairable item subject to failures governed by a \gls{nhpp} with instantaneous repair times. Specifically, we use the Power Law Intensity specification of the \gls{nhpp} due to its flexible polynomial failure intensity function, which can model both increasing and decreasing failure rates \cite{nist_hpp_model}. This function accommodates various failure patterns, including early-life failures, constant failure rates, and wear-out failures \cite{klutke2003critical}.  

For a \gls{nhpp} with failure intensity $h(t)$, the number of failures within any time interval \([t_{\text{agelt}}, t_{\text{age}}]\) follows a Poisson distribution \cite{potter2024bayesian,pishro2014introduction}:  
\begin{align}
    N(t_{\text{age}}) - N(t_{\text{agelt}}) \sim \text{Poisson}\left( \int_{t_{\text{agelt}}}^{t_\text{age}} h(\tau) d\tau \right),
\end{align}  
Here, \( N(t) \) is the \gls{nhpp} that counts the total number of failures of the hardware system up to time \( t \), \( t_{\text{age}} \) represents the system's current age, and \( t_{\text{agelt}} \) denotes the system's age at the latest repair.

The cumulative intensity function, \( H(t) \), is defined as:  
\begin{align}
    H(t) = \int_{0}^{t} h(\tau) d\tau,
\end{align}  
representing the expected number of failures in the interval \([0, t]\). Consequently, the probability of no failures in the interval \([t_{\text{agelt}}, t_{\text{age}}]\) is given by:  
\begin{align}
    R(t_\text{age}; t_\text{agelt}) &= P\left[N(t_{\text{age}}) - N(t_{\text{agelt}}) = 0 \right] \nonumber \\
    &= \exp{\left( -\int_{t_{\text{agelt}}}^{t_\text{age}} h(\tau) d\tau \right)},
    \label{eqn:survival}
\end{align}  
\cite{pishro2014introduction}. Similarly, the complementary probability of at least one failure in this interval, or otherwise interpreted as the probability of the next failure occuring before $t_\text{age}$, is
\begin{align}
    1 - R(t_{\text{age}}; t_\text{agelt}) = 1 - \exp{\left( -\int_{t_{\text{agelt}}}^{t_\text{age}} h(\tau) d\tau \right)}.
    \label{eqn:failure}
\end{align}

The interarrival times of failure events under a Power Law Intensity specification of the \gls{nhpp} follow a conditional Weibull distribution, characterized by the following intensity function:  
\begin{align}  
    h(t) = \alpha k t^{k-1}, \label{eqn:weibull_intensity}  
\end{align}  
and cumulative intensity:  
\begin{align}  
    H(t) = \alpha t^{k}, \label{eqn:weibull_cumulative}  
\end{align}  
where \(\alpha = \lambda^k\) and \(k\) are the rate raised to the power of the shape parameter and the shape parameter, respectively \cite{mcshane2008count,nist_hpp_model}.  

Using all the information above, the \gls{mle} of \gls{nhpp} parameters is obtained by solving the following optimization problem:
\begin{align}
    \hat{\alpha}, \hat{k} = 
    \argmin \prod_{j} \prod_{i} 
    & \left( 1 - R(t_{\text{age,ji}}; t_\text{agelt,ji}) \right)^{1-y_{ji}} \notag \\
    & \times  R(t_{\text{age,ji}}; t_\text{agelt,ji})^{y_{ji}} \label{eqn:likelihood},
\end{align}
where $j$ is the hardware system index, $i$ is the failure event index, and $y$ is the detection of a failure in the diagnostic test.

So far, the formulations have focused on modeling hardware system failures under a diagnostic test which fully covers the hardware system, otherwise known as a proof test. However, due to constraints like cost and time,  diagnostic tests which partially cover the hardware system are often used to gather partial knowledge of the system's operational status. This approach is particularly useful when certain subsystems of the hardware system exhibit higher failure rates, allowing for more cost-effective detection of failure modes \cite{jin2014reliability,torres2009modelling}. 

We directly account for diagnostic tests which partially test the hardware system by incorporating \gls{dc} in the inference of the reliability model parameters, which can improve accuracy by detailing failure modes and their allocation to subsystems \cite{forcina2020reliability}. The next subsection defines \gls{dc}, which is essential for formulating the reliability model likelihood function for parameter inference.

\subsection{Diagnostic Coverage}
\label{subsec:diagnostic_coverage}
\gls{dc} measures a diagnostic test's ability to detect and isolate failures, representing the percentage of potential failures that can be identified by the given diagnostic test\cite{SMITH2011103}. Reliability engineers use \gls{fmea}, expert judgement, or pre-defined failure rate tables such as Siemens SN 29500 database to determine the \gls{dc} of a hardware system \cite{stamatis2003failure,pepperl2020certification}. 

While \gls{dc} is traditionally defined using a time-independent failure intensity—often expressed as a constant average failure rate, we extend this to incorporate time-varying failure intensity. This approach reflects the evolving risk of failure over a hardware system's lifetime, where the failure intensity is time-dependent, representing the instantaneous failure rate at any given moment. A time-dependent intensity function is more appropriate for calculating \gls{dc}, as the effectiveness of a diagnostic test depends on when the test is performed during the hardware system's lifecycle. Thus, formally, \gls{dc} is defined as the ratio of detected failure intensity to total failure intensity:
\begin{align}
    c(t) = \frac{h_D(t)}{h_D(t)+h_{ND}(t)} \in [0, 1], \label{eqn:dc}
\end{align}
where $D$ stands for tested, $ND$ stands for not tested, and the total hardware system failure intensity is $h(t) = h_D(t)+h_{ND}(t)$.

To simplify reliability modeling of $B$ subsystems, we make the following assumptions:  

\begin{enumerate}  
    \item The hardware system is configured as a series of independent subsystems \cite{friedman1991reliability}.  
    \item Each subsystem follows a \gls{nhpp} with a common shape parameter \(k\), reflecting a similar life stage within the bathtub reliability curve.  
\end{enumerate}  

Building on these assumptions, we now show how subsystem model parameters can be reparameterized for $V$ diagnostic tests, which may overlap or form subsets within the \gls{dc}. Let us consider a hardware system consisting of $B$ subsystems
\[
S = \{s_1, s_2, \ldots, s_B\},
\]
where \( S_v \subseteq S \) represents the subsystems tested in diagnostic test \( v \). Let \( h_{s_i} \) denote the failure intensity of subsystem \( s_i \). The \gls{dc} equation for diagnostic test \( v \) is given as:
\begin{align}
    c_v(t) = \frac{\sum_{s_i \in S_v} h_{s_i}}{\sum_{s_i \in S} h_{s_i}}.
\end{align}

Assuming a common shape parameter \( k \) for all subsystems and that each subsystem follows a \gls{nhpp} with intensity function defined by \cref{eqn:weibull_intensity}, the \gls{dc} for test \( v \) simplifies to:
\begin{align}
    c_v = \frac{\sum_{s_i \in S_v} \alpha_{s_i}}{\sum_{s_i \in S} \alpha_{s_i}}. \label{eqn:dc_general}
\end{align}

The \gls{dc} in \cref{eqn:dc_general} is independent of time under the chosen \gls{nhpp} specification. However, the intensity functions \( h_{s_i}(t) \) remain time-dependent. 

Given the \glspl{dc} for the \( V \) diagnostic tests of the hardware system, we solve a system of \( V \) equations to express individual subsystem failure intensities \( h_{s_i} \) as functions of the total failure intensity and \glspl{dc}:
\begin{equation}
\begin{aligned}
    c_1 \sum_{s_i \in S} \alpha_{s_i} - \sum_{s_i \in S_1} \alpha_{s_i} &= 0, \\
    c_2 \sum_{s_i \in S} \alpha_{s_i} - \sum_{s_i \in S_2} \alpha_{s_i} &= 0, \\
    &\vdots \\
    c_V \sum_{s_i \in S} \alpha_{s_i} - \sum_{s_i \in S_V} \alpha_{s_i} &= 0.
\end{aligned}
\end{equation}

Then, the reliability of the subsystems tested with diagnostic test \( v \) is then given by:
\begin{align}
    P\left(\min(\{T_{s_i} : s_i \in S_v\}) > t \mid PT_1, \ldots, PT_V \right) \\
    = \prod_{v=1}^V \exp\left(-PT_v \int_{t_{agelt_v}}^{t} \alpha_v \tau^k\, d\tau\right).
\end{align}

For this paper, we focus on three types of diagnostic tests performed during maintenance: two diagnostic tests that have partial coverage (partial test 1 and partial test 2) of the hardware system, and a proof-test which has full coverage of the hardware system test. We explore two scenarios of redundant coverage by diagnostic tests: (1) overlapping diagnostic tests that examine some of the same subsystems, resulting in three subsystems (\cref{fig:blockvisualoverlap}), and (2) one diagnostic test being a subset of another diagnostic test, covering identical subsystems, also resulting in three subsystems (\cref{fig:blockvisualsubset}). The next subsections define the \gls{dc} for overlapping and subset diagnostic test scenarios.


\subsubsection{Overlapping Diagnostic Tests}
\label{subsubsec:overlappt}

\begin{figure}[h]
    \centering
    \includegraphics[width=0.5\linewidth]{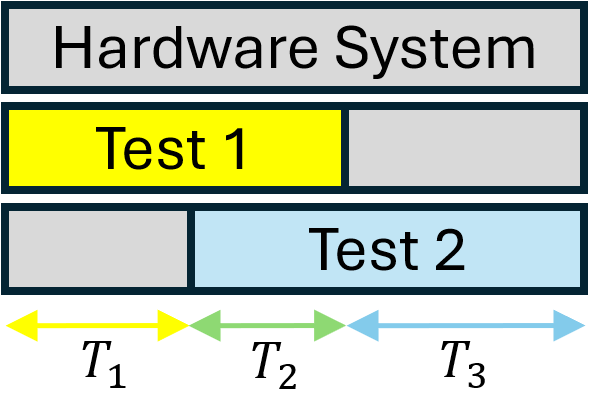}
    \caption{Visualization of overlapping diagnostic tests (partial test 1 and partial test 2), resulting in the hardware system being divided into three subsystems. The yellow is the subsystems tested by partial test 1, the blue is the subsystems tested by partial test 2, and the green color of the subsystem indicates which subsystem is redundantly tested in the diagnostic tests.}
    \label{fig:blockvisualoverlap}
\end{figure}

Let $T_1$ and $T_3$ denote the \gls{ttf} \gls{rv} of the subsystems included in partial test 1 but not partial test 2 and partial test 2 but not partial test 1, respectively (\cref{fig:blockvisualoverlap}). Further, let $T_2$ denote the \gls{ttf} \gls{rv} of the subsystems included in both partial test 1 and partial test 2.
Then, the \gls{ttf} \gls{rv} of the hardware system is expressed as:
\begin{align}
    T=\min(T_1,T_2,T_3) \label{eqn:ttf_subsystems}
\end{align}
\vspace{-1em}

Following the derivation process in \cref{subsec:diagnostic_coverage}, the \gls{dc} for partial test 1 and partial test 2 in overlapping diagnostic tests, respectively, is (see \cref{appsec:reliability_subset_derivation}):

\begin{align}
    c_1 &= \frac{\alpha_1 + \alpha_2}{\alpha_1 + \alpha_2 + \alpha_3}  \label{eqn:dc1},  \\
    c_2 &= \frac{\alpha_2 + \alpha_3}{\alpha_1 + \alpha_2 + \alpha_3} .\label{eqn:dc2}
\end{align}
where the \gls{dc} becomes a constant with respect to time due to the assumption of each subsystem having the same $k$ parameter from \cref{subsec:diagnostic_coverage} and with the constraints  $c_1+c_2 \geq 1$ and $0<c_1,c_2< 1$. Furthermore, the reparameterization of the subsystem \gls{nhpp} parameters in terms of the hardware system \gls{nhpp} parameters is 
\vspace{-1em}
\begin{align}
    \alpha_1  &=  \alpha \cdot (1-c_2), \\
    \alpha_2 &= \alpha \cdot (c_1+c_2 - 1), \\  
    \alpha_3 &= \alpha \cdot (1-c_1).
\end{align}

\subsubsection{Subset Diagnostic Tests}
\label{subsubsec:subsetpt}

\begin{figure}[h]
    \centering
    \includegraphics[width=0.5\linewidth]{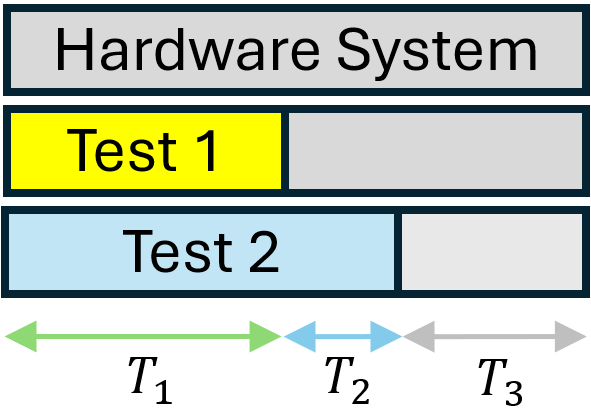}
    \caption{Visualization of subset diagnostic tests (partial test 1 and partial test 2), resulting in the hardware system being divided into three subsystems. The yellow is the subsystems tested by partial test 1, the blue is the subsystems tested by partial test 2, and the green color of the subsystem indicates which subsystem is redundantly tested in the diagnostic tests.}
    \label{fig:blockvisualsubset}
\end{figure}

For the subset diagnostic test configuration, the \gls{ttf} \gls{rv} of the hardware system is also given by \cref{eqn:ttf_subsystems}, with revised definitions for the \gls{ttf} \glspl{rv} \( T_1 \), \( T_2 \), and \( T_3 \). Let $T_1$ and $T_2$ denote the \gls{ttf} \gls{rv} of the subsystems included in partial test 1 but not partial test 2 and partial test 2 but not partial test 1, respectively (\cref{fig:blockvisualsubset}). Further, let $T_3$ denote the \gls{ttf} \gls{rv} of the subsystems not in both partial test 1 or partial test 2.  

As in \cref{subsubsec:overlappt}, we calculate the \gls{dc} for partial test 1 and partial test 2 in the subset diagnostic tests as follows:  

\begin{align}  
    c_1 &= \frac{\alpha_1}{\alpha_1 + \alpha_2 + \alpha_3}\\  
    c_2 &= \frac{\alpha_1 + \alpha_2}{\alpha_1 + \alpha_2 + \alpha_3}  
\end{align}  

Similarly, the same derivation process as \cref{subsubsec:overlappt}, but for subset of coverage in diagnostic tests, yields the following reparameterization of the subsystem \gls{nhpp} parameters in terms of the hardware system \gls{nhpp} parameters (see \cref{appsec:reliability_subset_derivation}): 



\begin{align}
    \alpha_1  &=  \alpha \cdot (c_1), \\
    \alpha_2 &= \alpha \cdot (c_2-c_1), \\  
    \alpha_3 &= \alpha \cdot (1-c_2).
\end{align}



During each maintenance cycle, partial test 1, partial test 2, or a proof-test is performed on the selected hardware systems, adding new labeled data to the dataset.

\subsection{Dataset Description}
\label{subsec:data}
At regular intervals of $\Delta t$, a maintenance cycle is performed on a subset of hardware systems, during which partial test 1, partial test 2, or a proof-test is conducted. Preventative maintenance is always performed during each diagnostic test, whether a failure is detected or not. Preventative maintenance is a standard maintenance practice that prevents unexpected failures, enhances safety, and extends the lifespan of hardware systems \cite{preventativemaintenance}. The dataset evolves over time and includes the diagnostic test results, ages of hardware systems, and ages of subsystems at the time of the diagnostic tests.

Formally, there are \( J \) hardware system instances in the population, and each instance of a hardware system has \( |I_j(t)| \) test results up to the given point in time $t$. Therefore, at a given point in time, the dataset contains: 
\[
M(t) = \sum_{j=1}^J|I_j(t)|
\]
test intervals (datapoints), and follows the relational form:
\begin{align*}
\mathcal{D}(t) = & \big(\mathbf{y} \in \{0,1\}^{M(t)}, \\ & \mathbf{t_{agelt1}}, \mathbf{t_{agelt2}}, \mathbf{t_{agelt3}},  \mathbf{t_{age}} \in \mathbb{R}^{M(t)} \big).
\end{align*}
Each binary element of the label vector \( \mathbf{y} \) represents a diagnostic test result, where 1 indicates a failure detected and 0 indicates no failure detected. The elements of \( \mathbf{t_{age}} \), \( \mathbf{t_{agelt1}} \), \( \mathbf{t_{agelt2}} \), and \( \mathbf{t_{agelt3}} \) correspond to instances of the hardware system age at test interval \( i \), and the ages at the most recent tests of subsystems 1, 2, and 3, respectively. For each hardware system $j$, only one type of diagnostic test is performed at a given maintenance cycle. Throughout the manuscript, we drop the dependency of $\mathcal{D}$ and $M$ on $t$ for cleaner notation.

\begin{figure}
    \centering
    \includegraphics[width=0.95\linewidth]{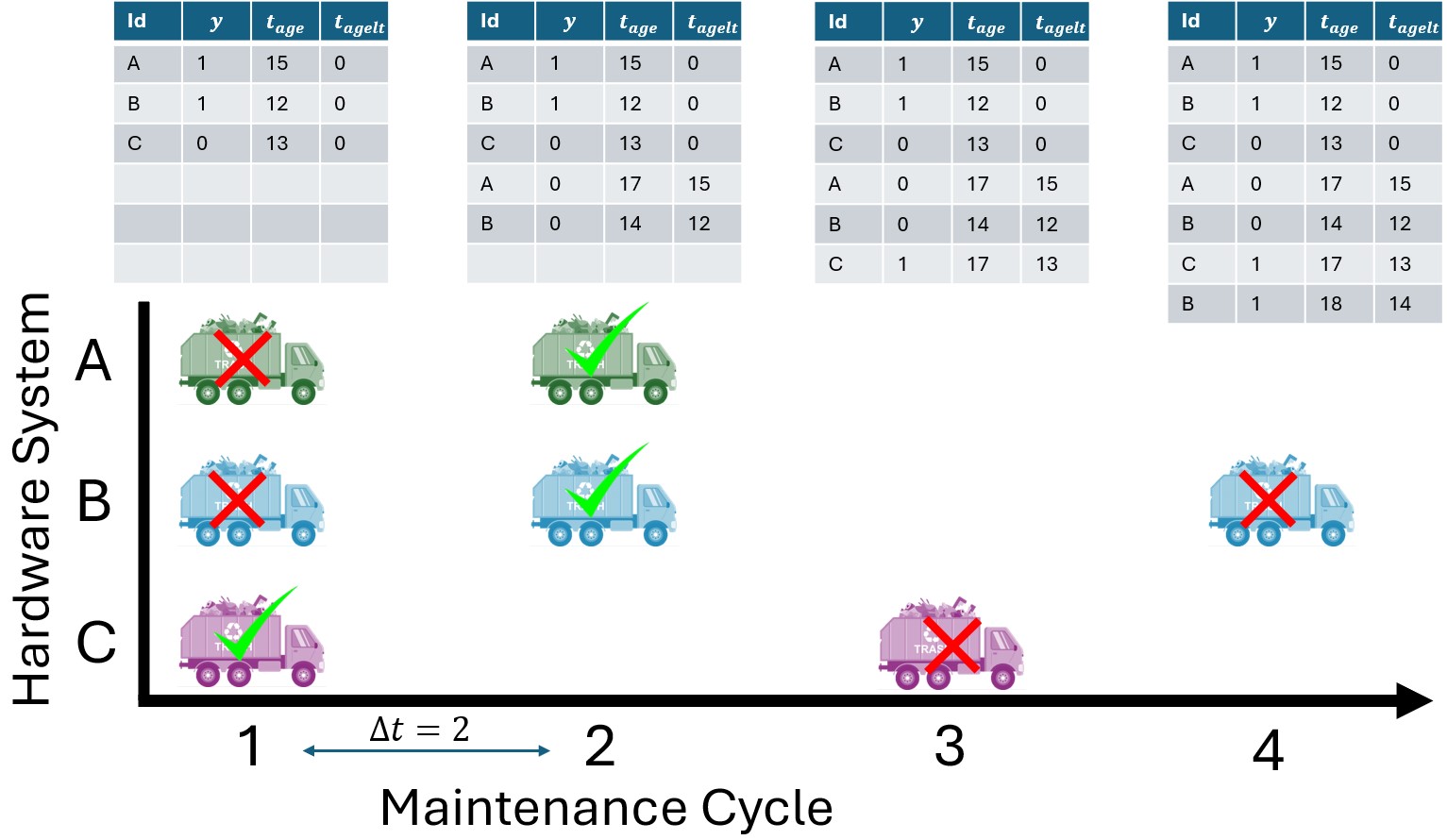}
    \caption{An example of how the dataset $\mathcal{D}(t)$ evolves over four maintenance cycles is provided. Consider three hardware system instances ($J=3$), with maintenance cycles occurring at intervals of $\Delta t = 2$ months. Over the four maintenance cycles, the dataset sizes are $M=3$, $M=5$, $M=6$, and $M=7$. }
    \label{fig:dataset_example}
\end{figure}

For example, consider diagnostic tests in \cref{fig:dataset_example}, where hardware systems A, B, and C undergo 4 maintenance cycles of proof testing. The dataset evolves depending on which systems are tested in each maintenance cycle, with each cycle updating the tested systems’ ages and adding new data. The dataset at each cycle is shown, with the number of test intervals M increasing over time. The dataset $D(t)$ used for reliability model parameter inference is shown at each maintenance cycle in \cref{fig:dataset_example}.

The following sections present the reliability model likelihood when diagnostic tests are partial test 1, partial test 2, and the proof-test, along with our proposed \gls{al} \gls{af} for optimizing diagnostic test and hardware system instance selection.


\vspace{-1em}
\section{Methodology}
\label{sec:methodology}

The following subsections explain how \gls{dc} is integrated into the likelihood equation for modeling failure data of repairable hardware systems and outlines a proposed optimization strategy for selecting diagnostic tests at a given maintenance cycle. Overall, the \gls{al} cycle for repairable hardware systems with partial diagnostic tests is shown in \cref{fig:cycleofreliability}.

We make the following simplifying assumptions:
\begin{itemize}
    \item  DC is correct: We assume the DC is accurately specified using techniques like \gls{fmea} or expert judgment \cite{stamatis2003failure, pepperl2020certification}.
    \item Model Mismatch: We assume no mismatch between the data-generating process and the reliability model, both following a NHPP with a Power Law Intensity. 
\end{itemize}

\subsection{Reliability Modeling with Partial Testing}

Leveraging \gls{dc} and the impact of partial tests and proof-tests on hardware system failure intensity, we update the likelihood and reliability models in \cref{subsec:failure_modeling} to improve parameter estimation, reflecting real-world testing processes \cite{maranzanosubystem}. Notably, although only a portion of the hardware system is tested during partial tests, the information gained enhances our understanding of the entire system, as subsystem failure intensity is expressed relative to the total hardware system failure intensity.

\begin{figure}[h]
    \centering
    \includegraphics[width=0.95\linewidth]{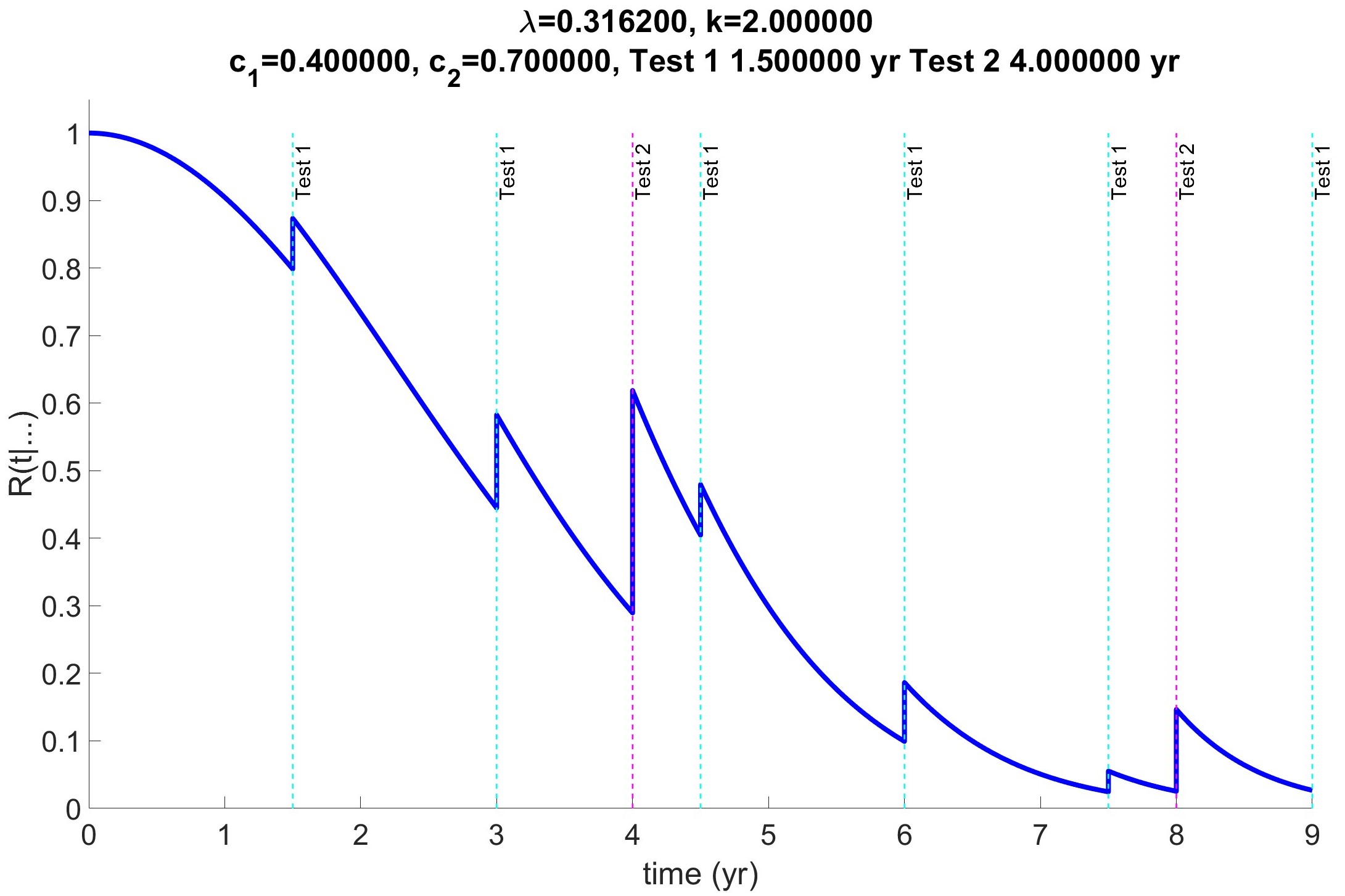}
    \caption{Example of the reliability of a hardware system with diagnostic tests partial test 1 and partial test 2 during maintenance cycles for subset configurations.
}
    \label{fig:reliability_curve}
\end{figure}

\textbf{Overlapping Diagnostic Tests: }
The reliability \( R(t_\text{age}) \) of an instance of the hardware system varies depending on the specific diagnostic test performed during the given maintenance cycle:
\begin{align}
R(t|\vec{PT}) =& \exp \bigr(- \alpha \times PT_1 (1-c_2) (t_{age}^{k} -t_{agelt1}^{k}) \bigr) \times  \nonumber\\
& \exp \bigr(-\alpha  \times PT_2 (c_2+c_1-1) (t_{age}^{k} - t_{agelt}^{k}) \bigr) \times \nonumber\\
& \exp \bigr(-\alpha \times PT_3 (1-c_1) (t_{age}^k - t_{agelt3}^k) \bigr) \label{eqn:overlap_reliability}
\end{align}
where $\vec{PT}=[PT_1,PT_2,PT_3]$. The derivations are given in \cref{appsec:reliability_derivation}. An example of the reliability curve specified by \cref{eqn:overlap_reliability} over the history of several diagnostic tests is shown in \cref{fig:reliability_curve}, revealing the emergence of a complex ``sawtooth'' pattern. 




\textbf{Subset Diagnostic Testing: }
Similar to \cref{eqn:overlap_reliability}, the reliability equation $R(t_\text{age})$ of the instance of the hardware system varies depending on the specific diagnostic test performed during the given maintenance cycle:
\begin{align}
R(t|\vec{PT}) = &\exp \bigr(- \alpha \times PT_1 c_1 (t_{age}^{k} -t_{agelt}^{k}) \bigr) \times  \nonumber\\
& \exp \bigr(-\alpha  \times PT_2 (c_2-c_1) (t_{age}^{k} - t_{agelt2}^{k}) \bigr) \times \nonumber\\
& \exp \bigr(-\alpha \times PT_3 (1-c_2) (t_{age}^k - t_{agelt3}^k) \bigr) \label{eqn:subset_reliability}
\end{align}
where the derivations are given in \cref{appsec:reliability_subset_derivation}.


The likelihood function in \cref{eqn:likelihood} is updated by replacing \( R(t_\text{age}; t_\text{agelt}) \) with either \cref{eqn:overlap_reliability} or \cref{eqn:subset_reliability}, depending on whether the diagnostic tests have subset or overlap in coverage. Furthermore, depending on the diagnostic test performed and the configuration, $PT_1$,$PT_2$ and $PT_3$ are assigned different values from ${0,1}$ following \cref{tab:partialtestassign}.

\begin{table}[h!]
\centering
\begin{tabular}{c|lcc}
\toprule
           & \multicolumn{2}{c}{$[PT_1, PT_2, PT_3]$} \\ 
\cmidrule(lr){2-3}
           Diagnostic Test & Overlapping & Subset  \\ 
\midrule
partial test 1 & $[1, 1, 0]$ & $[1, 0, 0]$  \\ 
partial test 2 & $[0, 1, 1]$ & $[1, 1, 0]$ \\ 
proof test & $[1, 1, 1]$ & $[1, 1, 1]$ \\ 
\bottomrule
\end{tabular}
\caption{Assignment of variables $[PT_1, PT_2, PT_3]$ depending on the diagnostic test and configuration. \( 1 \) indicates that a subsystem of the hardware system underwent the diagnostic test and \( 0 \) otherwise.}
\label{tab:partialtestassign}
\end{table}

Subsystem-level diagnostic tests offer a cost-efficient alternative to full testing, achieving the same \gls{mse} on reliability model parameters under a desired error threshold \cite{maranzanomaximum,mcshane2008count}. To capitalize on this efficiency, we aim to optimize diagnostic test selection by maximizing the trace of the \gls{fim} under budget constraints while leveraging \gls{dc}.

\subsection{Optimization Strategy}
\label{subsec:proposedstrategy}
We employ the \gls{fim} to quantify the expected information gain about the unknown model parameters based on the anticipated diagnostic test results. To motivate this, the expected \gls{kld} between the likelihood evaluated at the estimated and true model parameters, with respect to $\mathbf{q}$, is asymptotically proportional to the dataset size as:
\begin{align}
    \propto \text{tr}(\mathbf{A}_q(\mathbf{\theta})^{-1} \mathbf{A}_p(\mathbf{\theta})) \label{eqn:fim_var_reduction}
\end{align}
where tr denotes trace, and \( \mathbf{A}_q \) and \( \mathbf{A}_p \) are the \gls{fim} with respect to the assumed training data-generating distribution $\mathbf{q}$ and the true data-generating distribution $\mathbf{p}$, respectively. $\mathbf{q}$ represents the assumed distribution over the "training data" used for model inference, and $\mathbf{p}$ is the true distribution over the "testing data." Here, $\mathbf{\theta}$ denotes the true model parameters \cite{zhang2000value}. Minimizing the \gls{kld} means the estimated model parameters are ``close'' to the true model parameters.

The trace inequality applied to \cref{eqn:fim_var_reduction} yields the following (assuming positive definiteness for \glspl{fim}):
\begin{align}
    \text{tr}(\mathbf{A}_q(\mathbf{\theta})^{-1} \mathbf{A}_p(\mathbf{\theta})) \leq \text{tr}(\mathbf{A}_q(\mathbf{\theta})^{-1}) \text{tr}( \mathbf{A}_p(\mathbf{\theta})) \label{eqn:traceineq}
\end{align} 
When optimizing for the training distribution $\mathbf{q}$, as \( \text{tr}( \mathbf{A}_p(\mathbf{\theta})) \) is constant with respect to $\mathbf{q}$, the upper-bound minimization becomes equivalent to \cite{sourati2017probabilistic}:
\begin{align}
    \argmin_{q} \text{tr}(\mathbf{A}_q(\mathbf{\theta})^{-1}) \label{eqn:minvar}
\end{align}

Optimizing the upper bound is numerically more efficient, as it avoids matrix multiplication and accounts for the potential lack of knowledge of the test distribution $p$.
Minimizing \cref{eqn:minvar} also reduces the expected \gls{kld}. In \cref{eqn:minvar}, optimizing the training data distribution $\mathbf{q}$  minimizes the variance of model parameter estimates. Given the optimal $\mathbf{q}$ from \cref{eqn:minvar}, sampling from $\mathbf{q}$ selects the optimal data points for oracle labeling, aiding model parameter inference
\cite{sourati2016mutual,sourati2017probabilistic}.

\begin{table*}[h!]
    \centering
    \caption{Parameter Settings for Different Test Styles}
    \label{tab:test_styles}
    \begin{tabular}{p{3cm}cccccccccccccccc}
        \toprule
        \textbf{Test Scenario} & \textbf{\gls{dc}} & 1 & 2 & 3 & 4 & 5 & 6 & 7 & 8 & 9 & 10 & 11 & 12 & 13 \\ 
        \midrule
        \multirow{2}{*}{\textbf{Overlapping}} 
        & $c_1$ & 0.3 & 0.3 & 0.4 & 0.5 & 0.5 & 0.5 & 0.6 & 0.6 & 0.6 & 0.7 & 0.7 & 0.8 & - \\ 
        & $c_2$ & 0.8 & 0.9 & 0.9 & 0.6 & 0.8 & 0.9 & 0.7 & 0.8 & 0.9 & 0.8 & 0.9 & 0.9 & - \\ 
        \midrule
        \multirow{2}{*}{\textbf{Subset}} 
        & $c_1$ & 0.1 & 0.1 & 0.1 & 0.2 & 0.2 & 0.2 & 0.3 & 0.3 & 0.3 & 0.4 & 0.4 & 0.5 & 0.5 \\ 
        & $c_2$ & 0.8 & 0.7 & 0.6 & 0.8 & 0.7 & 0.6 & 0.8 & 0.7 & 0.6 & 0.8 & 0.7 & 0.8 & 0.9 \\ 
        \bottomrule
    \end{tabular}
\end{table*}

Building on this, \cite{sourati2017probabilistic} proposed the following \gls{sdp} as the \gls{al} \gls{af} strategy:
\begin{align}
&\argmin_{\mathbf{v}_1, \mathbf{v}_2, \ldots, \mathbf{v}_d, \mathbf{q}_1, \ldots, \mathbf{q}_M} \mathbf{v}_1 + \ldots + \mathbf{v}_d \label{eqn:souratisdp}\\
    &\text{subject to} \bigoplus_{k=1}^{d} \begin{bmatrix}
        \sum_{j=1}^{J} \mathbf{q}_j \mathbf{A}_j &  \mathbf{e}_k \\ \mathbf{e}_k^T & \mathbf{v}_k 
    \end{bmatrix} \succ 0 \nonumber 
\end{align}
where $\mathbf{A}$ is the conditional \gls{fim} of the unlabeled data pool and $\mathbf{e}_k$ is a unit vector with all zeros except for a one in the $k$-th position. Sampling from the optimal $\mathbf{q}$ given by this SDP selects the optimal data points for oracle labeling, balancing exploration and exploitation.

 The \gls{al} \gls{af} \cref{eqn:souratisdp} is enhanced to address aging hardware systems undergoing preventative maintenance. Specifically, we account for partial coverage of the diagnostic tests,  diagnotic testing budget constraints,  and the hardwaare system aging, all of which affect reliability during a given maintenance cycle. The resulting \gls{al} \gls{af} is a relaxed \gls{misdp}:

\begin{align}
&\argmin_{\mathbf{v}_1,\mathbf{v}_2,\ldots,\mathbf{v}_d,q_{11},\ldots,q_{1V},\ldots,q_{J1},\ldots,q_{JV}} \mathbf{v}_1 + \ldots + \mathbf{v}_d \\
    &\text{such that} \qquad \sum_{j=1}^{J} \sum_{i=1}^{V} w_{ji} q_{ji} \leq B \nonumber \\
    & \qquad \bigoplus_{k=1}^{d} \begin{bmatrix}
        \sum_{j=1}^{J} \sum_{i=1}^{V} q_{ji} \mathbf{A}_{ji} &  \mathbf{e}_k \\ \mathbf{e}_k^T & \mathbf{v}_k 
    \end{bmatrix} \succ 0 \nonumber \\ 
    & \qquad \sum_{i=1}^{V} {q_{ji}} \leq 1 ~ ; ~ q_{j.} \geq 0 ~ ; ~ q_{j.} \leq 1 ; \qquad \forall j\nonumber \\
\end{align}
where $\mathbf{A}_{ji}$ is the \gls{fim} based on the $j$ hardware instance's $i$ diagnostic test,  $V$ is the number of diagnostic test options available (in this case $V=3$), \( w_{ji} \) represents the cost of performing diagnostic test \( i \) on hardware system instance \( j \), and \( q_{ji} \) acts as a soft selection indicator for performing diagnostic test \( i \) on hardware system instance \( j \). 

\section{Experiments}

The time-to-failure data for hardware systems was simulated using a conditional Weibull distribution. We conduct 100 \gls{mc} trials for both the overlapping coverage and subset coverage scenarios under three parameter settings for $\alpha$ and $k$: (0.1, 1.3), (0.5, 0.5), and (0.25, 2). These settings correspond to different life stages of the hardware system as described by the reliability bathtub curve  \cite{OHRING1995747}: infant mortality (decreasing failure intensity), useful life (approximately constant failure rate), and end-of-life wear-out (increasing failure intensity). We evaluate all permutations of the parameter settings in combination with the permutations from \cref{tab:experiment_params} and \cref{tab:test_styles}. The goal is to assess the effectiveness of our proposed \gls{al} \gls{af} in selecting diagnostic tests and hardware system instances under budget constraints. More details on the simulation of maintenance cycles,  hardware system failures, and dataset generation are provided in \cref{algo:al_overlap_coverage}. Failure data for the hardware system were generated using a conditional Weibull distribution. 

\begin{table}[h!]
    \centering
    \caption{Parameter Settings for Experimentation}
    \label{tab:experiment_params}
    \begin{tabular}{p{3cm}|p{5cm}}
        \hline
        \textbf{Parameter} & \textbf{Values} \\ \hline
        $J$ & [50, 100] \\
        Budgets & [5, 10, 25] \\
        \gls{al} Acquisition Function & [random, oldest, likely failure, entropy, ours] \\
        $\Delta t$ & [2.5, 5] \\ \hline
    \end{tabular}
\end{table}

\subsection{Hardware}
All experiments were conducted on a system with Dual Intel Xeon E5-2650 processors running at 2 GHz, equipped with 32 cores and 16 GB of RAM. 

\begin{algorithm*}
\label{algo:al_overlap_coverage}
\caption{Active Learning with Overlapping Coverage and Budget Constraints}
\begin{algorithmic}[1]
\Statex \textbf{Input:} Acquisition Function $a$, Diagnostic Test costs $b_1, b_2, b_3$, Budget $B$, \gls{dc} Coefs $c_1, c_2$, Model Parameters $\alpha,k$
\State Init Labeled Set $\mathcal{L}$, Unlabeled Set $\mathcal{U}$, Failure Time Set $\Gamma$, model $R(t|\hat{\alpha}, \hat{k})$
\For{$\text{maintenance cycle} = 1 ~\text{to} ~ M$}
    \State $\mathcal{S} \gets a(\mathcal{U},\mathcal{T},R,b_1,b_2,b_3,B)$ \Comment{scores for instances $\mathcal{I}$, tests $\mathcal{T}$ using $a$}
    \State $\mathcal{I}_\text{selected}, \mathcal{T}_\text{selected} \gets$ \textsc{budget\_constraint}($\mathcal{S}$, $\mathcal{I}$, $\mathcal{T}$, $b_1, b_2, b_3$, $B$) \Comment{Apply budget constraints}
    \State $\mathcal{L}, \mathcal{U} \gets$ \textsc{maintenance\_check}($\mathcal{L}$, $\mathcal{U}, \mathcal{I}_\text{selected}$, $\mathcal{T}_\text{selected}$, $\Gamma$, $c_1, c_2$) \Comment{perform selected diagnostic tests and maintenance}
    \State Fit model $R(t|\hat{\alpha}, \hat{k})$ with labeled set $\mathcal{L}$
\EndFor
\end{algorithmic}
\end{algorithm*}

\begin{algorithm*}
\caption{Update for Labeled and Unlabeled Population in Overlapping Configuration}
\begin{algorithmic}[1]
\Function{maintenance\_check\_update}{labeled dataset $\mathcal{L}$, unlabeled dataset $\mathcal{U}$, instances $\mathcal{I}$, tests $\mathcal{T}$, failure times $\Gamma$, diagnostic coefficients $c_1, c_2$}
    \Statex
    \For{each instance $i \in \mathcal{I}$} \Comment{Update the labeled set $\mathcal{L}$}
        \State Perform diagnostic test $\mathcal{T}_i$ and receive failure detection $y_{i}$
        \State Retrieve  $t_{\text{agelt1},i},t_{\text{agelt2},i},t_{\text{agelt3},i}$  from $\mathcal{U}$ \Comment{Ages of subsystems at the time of their last diagnostic test.}
        \State Retrieve $t_{\text{age},i}$ from $\mathcal{U}$ \Comment{Age of hardware system at time of diagnostic test}
        \State Set $PT_{1},PT_{2},PT_3$ based on diagnostic test $\mathcal{T}_i$
        \State $\mathcal{L} \gets \mathcal{L} \cup \{ (y_{i},t_{\text{age},i}, t_{\text{agelt1},i},t_{\text{agelt2},i},t_{\text{agelt3},i},PT_1,PT_2,PT_3) \}$ \Comment{Update the labeled set}
    \EndFor
    \Statex
    \State Calculate $\alpha_1,\alpha_2, \alpha_3$ based on \gls{dc} coefs and $\alpha$
    \For{each instance $i \in \mathcal{I}$} \Comment{Update the unlabeled set $\mathcal{U}$}
        \State Retrieve age $t_{\text{age},i}$ at time of diagnostic test from $\mathcal{U}$
        \State Retrieve partial test ages: $t_{\text{agelt1},i}, t_{\text{agelt2},i}, t_{\text{agelt3},i}$ from $\mathcal{U}$
        \If{$\mathcal{T}_i = 1$}
            \State $t_{\text{agelt1},i}, t_{\text{agelt2},i} \gets t_{\text{age},i}$
            \State $\Gamma[i, 0] \gets \text{WeibullGenerator}(\alpha_1, k, t_{\text{age},i})$
            \State $\Gamma[i, 1] \gets \text{WeibullGenerator}(\alpha_2, k, t_{\text{age},i})$  \Comment{Regenerate age at failure times for subsystems 1 and 2}
            
        \ElsIf{$\mathcal{T}_i = 2$}
            \State $t_{\text{agelt2},i}, t_{\text{agelt3},i} \gets t_{\text{age},i}$
            \State  $\Gamma[i, 1] \gets \text{WeibullGenerator}(\alpha_2, k, t_{\text{age},i})$
            \State  $\Gamma[i, 2] \gets \text{WeibullGenerator}(\alpha_3, k, t_{\text{age},i})$ \Comment{Regenerate age at failure times for subsystems 2 and 3}
        \ElsIf{$\mathcal{T}_i = 3$}
            \State $t_{\text{agelt1},i}, t_{\text{agelt2},i}, t_{\text{agelt3},i} \gets t_{\text{age},i}$
            \State $\Gamma[i, 0] \gets \text{WeibullGenerator}(\alpha_1, k, t_{\text{age},i})$
            \State $\Gamma[i, 1] \gets \text{WeibullGenerator}(\alpha_2, k, t_{\text{age},i})$
            \State $\Gamma[i, 2] \gets \text{WeibullGenerator}(\alpha_3, k, t_{\text{age},i})$  \Comment{Regenerate age at failure times for subsystems 1, 2 and 3}
        \EndIf
        \State Increment age: $t_{\text{age},i} \gets t_{\text{age},i} + \Delta t$
        \State Update $\mathcal{U}[i] \gets [t_{\text{age},i}, t_{\text{agelt1},i}, t_{\text{agelt2},i}, t_{\text{agelt3},i}]$
    \EndFor

    \State For instances not tested, increment age by $\Delta t$
    \State \Return updated labeled set $\mathcal{L}$, unlabeled set $\mathcal{U}$
\EndFunction
\end{algorithmic}
\begin{algorithmic}[1]
\Function{WeibullGenerator}{ $\alpha$, $k$,$t_p$}
    \State $u \gets \text{Random}()$ \Comment{Generate a random number $u$ in the range [0, 1]}
    \State $t_n \gets \left( \frac{\ln(1 - u)}{-\alpha} + t_p^k \right)^{\frac{1}{k}}$
    \State \Return $t_n$
\EndFunction
\end{algorithmic}

\end{algorithm*}

\subsection{Active Learning Acquisition Functions}
In these experiments, the \gls{al} \glspl{af} $a$ determine both which diagnostic test—partial test 1, partial test 2, or a proof test—to perform and the hardware system instance on which to perform it during each maintenance cycle. Notably, each diagnostic test is tailored to detect failures in specific subsystem combinations rather than addressing the hardware system as a whole. To assess the effectiveness of our proposed \gls{al} \gls{af} we compare it against widely used \gls{al} strategies from the literature, as well as intuitive \gls{al} strategies specifically designed for reliability model parameter inference. The \gls{al} \glspl{af} used for comparison are as follows: 
\\ \\
\noindent \textbf{Oldest Subsystem}: select the diagnostic test and hardware system instance combination that targets the subsystem with the longest time since its last test.
    \[
    a_{\text{oldest}} = \argmin_{i \in \{1,2,3\}, j\in J} \, t_{\text{agelt,ji}}
    \]

\noindent \textbf{Most Likely Failure}: select the instance of the hardware system with the highest failure probability at the time of the maintenance cycle
    \[
    a_{\text{likely failure}} = \argmax_{i \in \{1,2,3\}, j \in J} \, 1-R(t_\text{age,ji}|PT_1,PT_2,PT_3)
    \]

\noindent \textbf{Entropy}: select the combination of diagnostic test and hardware system instance with highest entropy at the time of the maintenance cycle
    \[
    \begin{aligned}
     a_{\text{entropy}} = &\argmax_{i \in \{1,2,3\}, j \in J} \left[ -R(t_\text{age,ji}|\vec{PT}) \ln{R(t_\text{age,ji}|\vec{PT})} \right. \\
    & \left. - (1 - R(t_\text{age,ji}|\vec{PT})) \ln{(1 - R(t_\text{age,ji}|\vec{PT}))} \right]
    \end{aligned}
    \]

\noindent \textbf{Random}: uniformly at random select the diagnostic test and the instance of the hardware system.


\begin{figure*}[h!]
    \centering
    \begin{subfigure}[t]{0.8\textwidth}
        \centering
        \includegraphics[width=\textwidth]{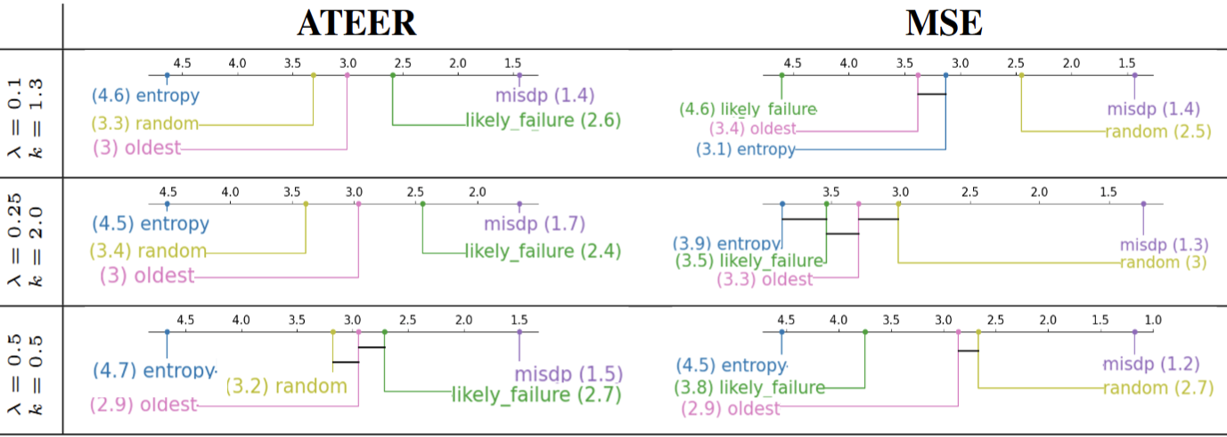}
        \caption{Overlapping configuration}
        \label{fig:cdoverlapping}
    \end{subfigure}
    \hfill
    \begin{subfigure}[t]{0.8\textwidth}
        \centering
        \includegraphics[width=\textwidth]{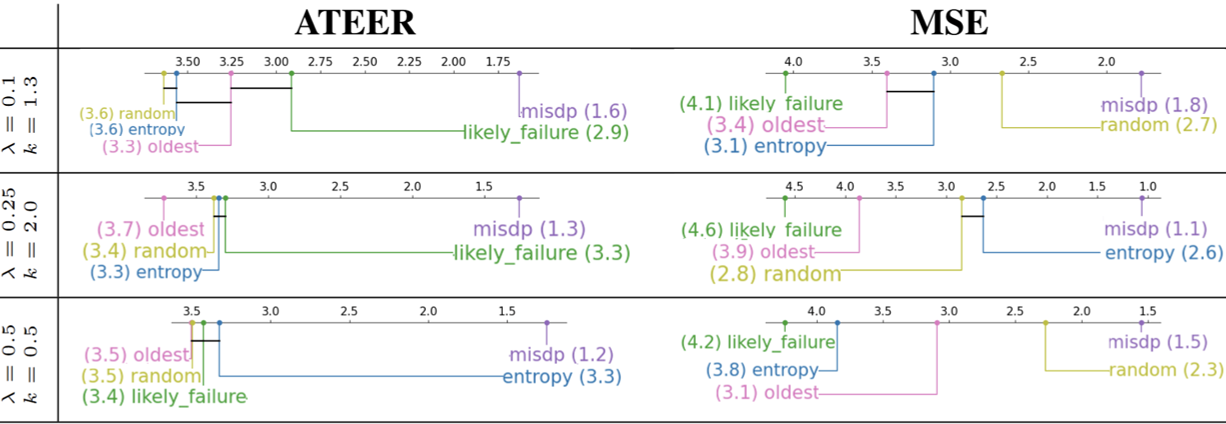}
        \caption{Subset configuration}
        \label{fig:cdsubset}
    \end{subfigure}
    \caption{Critical difference diagrams for acquisition functions across experiment realizations. The critical difference diagrams display the average ranks of the acquisition functions across various experiment realizations, considering combinations of diagnostic coverage coefficients, maintenance cycle frequency, and maintenance budget. For both \gls{ateer} and \gls{mse}, a lower rank indicates a better acquisition function.}
    \label{fig:cd_both}
\end{figure*}

\subsection{Metrics}
To quantify the performance of the \gls{al} \glspl{af}, we compute the \gls{auc} of the \gls{alc}, where the points comprising the \gls{alc} are derived from standard reliability analysis metrics, as detailed below. We compute the \gls{auc} of the error metrics with respect to the number of maintenance cycles. At each maintenance cycle cycle, we evaluate the following error metrics: \gls{ateer} and \gls{mse}.

\begin{figure}[h] \centering \includegraphics[width=0.95 \linewidth]{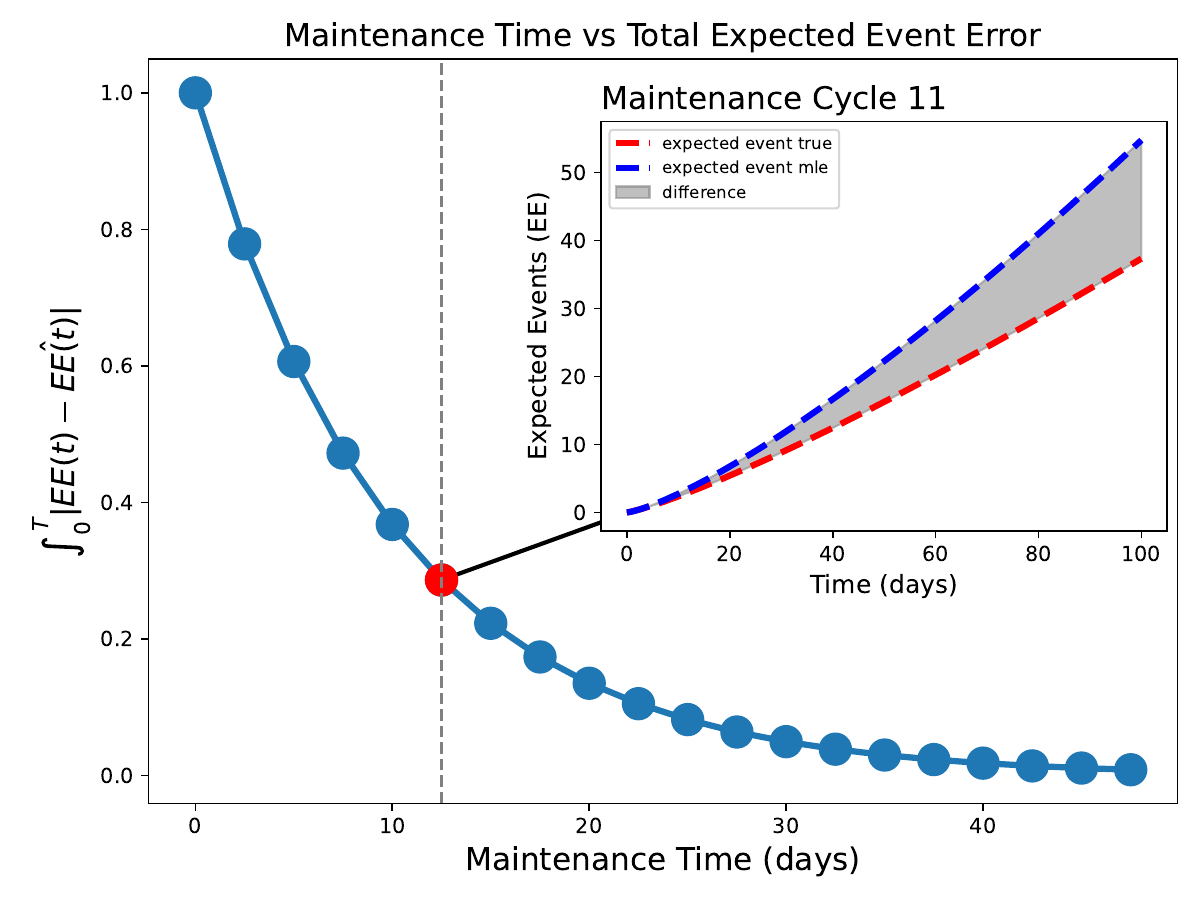} \caption{Example illustrating the \gls{ateer} metric.} \label{fig:expected_event_example} \end{figure}

\textbf{\gls{ateer}}: For \gls{nhpp}, the cumulative intensity fully characterizes the stochastic process, capturing both the inter-arrival distributions and statistical moments of the process within a given time interval. Therefore, for each maintenance cycle, we compute the \gls{ateer} using the current estimate of the model parameters and the true model parameters that generate the data:

\begin{align}
\int_{t=0}^T \left| \hat{\alpha} \tau^{\hat{k}} - \alpha \tau^k \right| d\tau \label{eqn:ateer}
\end{align}
\cref{eqn:ateer} represents a point on the \gls{ateer} \gls{alc} for a given maintenance cycle (see \cref{fig:expected_event_example}).
Since the total error in the expected event curves does not typically converge towards zero as $T \rightarrow \infty$, we use a large value of $T = 100$ in practice. 

\begin{figure*}
    \centering
    \includegraphics[width=0.98\linewidth]{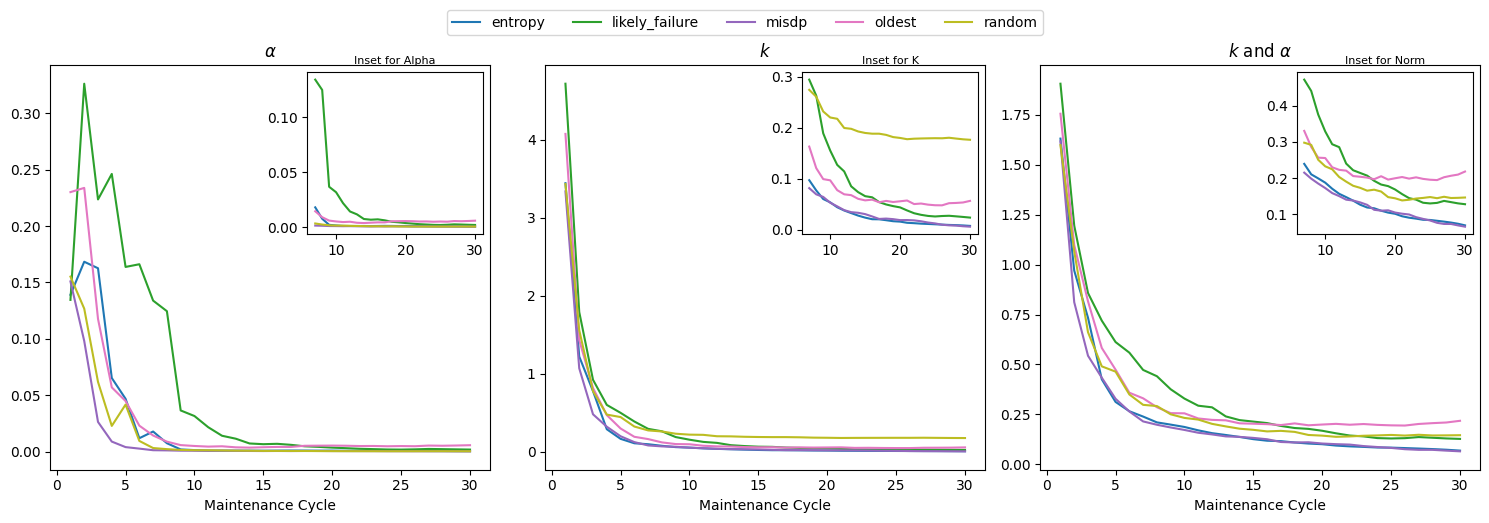}
    \caption{\gls{mse} \gls{alc} for each individual reliability model parameter, $\alpha,k$ for the experiment of $k=1.3,\lambda=0.1,c_1=0.2, c_2=0.6, \Delta t = 5.0, J=50, B=5.0$ with the subset configuration.}
    \label{fig:example_mse}
\end{figure*}

\textbf{\gls{mse}}: \gls{mse} is commonly used as an evaluation metric in \gls{al} as it is equivalent to the trace of the \gls{fim} for an unbiased estimator of the model parameters. \gls{mse} is calculated between the estimated parameter vector $[\hat{\alpha} , \hat{k}]$ and the true parameter vector $[\alpha, k]$ at a given maintenance cycle.

\subsection{Results}
\label{subsec:results}

We demonstrate, through simulation, that our proposed \gls{al} \gls{af} consistently ranks highest on average across all stages of the hardware system lifecycle (infant mortality, useful life, and end-of-life), outperforming common \glspl{af} like entropy and more intuitive methods such as the \textit{Most Likely Failure} (\cref{fig:cdsubset,fig:cdoverlapping}). 

To validate the superior performance of our proposed \gls{al} \gls{af},  
we conduct a Friedman hypothesis test at a significance level of 0.05. This non-parametric test is well-suited for repeated measurements, treating the \gls{al} \glspl{af} (random, oldest subsystem, most likely failure, entropy, ours) as ``treatments'' and the combinations of budget, \gls{dc} values, and maintenance cycle frequency ($\Delta t$) as ``blocks.'' The Friedman test evaluates whether there are statistically significant differences in the ranks of the \gls{al} \glspl{af} across a wide range of simulation scenarios.

The Friedman test revealed significant differences among the ranks of the \gls{al} \glspl{af}. Consequently, we performed a post-hoc Conover-Iman test to analyze pairwise differences. To control the false discovery rate resulting from repeated pairwise tests, we applied the Benjamini-Hochberg procedure for p-value adjustment. This ensured rigorous control over the risk of false positives. Our results confirm that our proposed \gls{al} \gls{af} consistently achieves statistically significant improvements over the other methods. The findings are visualized in critical difference diagrams (\cref{fig:cdoverlapping,fig:cdsubset}), illustrating performance in the overlapping and subset scenarios, respectively. 


\section{Discussion}
\label{sec:discussion}
We observe that our \gls{al} \gls{af} on average has significantly lower \gls{ateer} or \gls{mse} over the first few maintenance cycles, and most of the methods converge to the correct reliability model parameters after 20 maintenance cycles. However, our method rapidly reduces the \gls{ateer} or \gls{mse} during the first few maintenance cycles, as the \gls{al} \gls{af} is specifically designed to select diagnostic tests and hardware instance combinations that minimize the \gls{kld} between the data-generating distribution with the true reliability model parameters and the distribution with the estimated reliability model parameters. 

\begin{figure}[h!]
    \centering
    \includegraphics[width=1.03\linewidth]{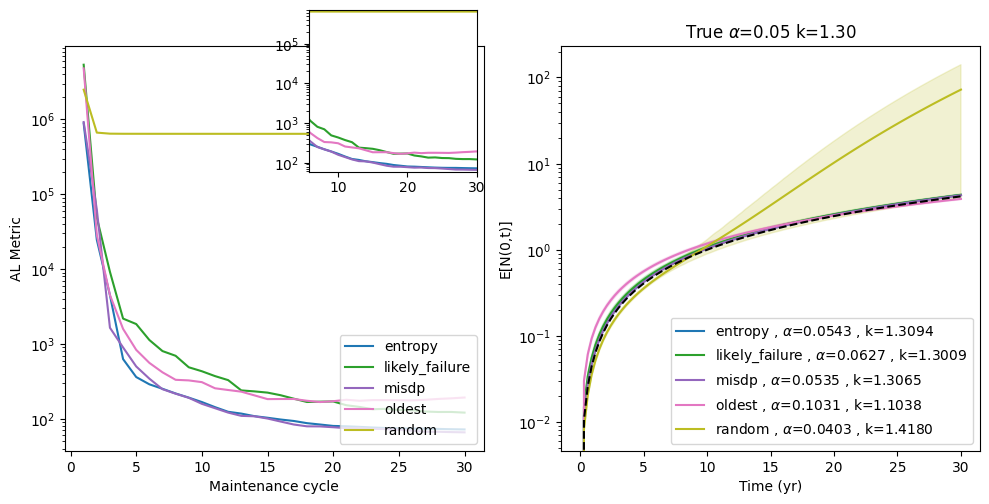}
    \caption{The left figure shows the \gls{ateer} \gls{alc}, while the right figure illustrates the expected number of events by time $t$ for the final reliability model parameters, after all maintenance cycles, under the experiment settings of $k = 1.3$, $\lambda = 0.1$, $c_1 = 0.2$, $c_2 = 0.6$, $\Delta t = 5.0$, $J = 50$, and $B = 5.0$ in the subset configuration.}
    \label{fig:example_ee}
\end{figure}

For the \text{Oldest Subsystem} \gls{af}, in the first maintenance cycle, all hardware systems are manufactured around the same time, causing the \gls{af} to behave similarly to random selection of the diagnostic test and hardware instance combination, which is suboptimal. For the \text{Most Likely Failure} \gls{af}, in the early maintenance cycles, the estimated reliability model parameters deviate significantly from the true parameters. As a result, the hardware instances deemed most likely to fail at a given maintenance cycle may actually not fail with high probability.  The \text{entropy} \gls{af} is less effective for reliability model parameter inference because this problem does not directly align with binary classification. In other words, there is no ``decision boundary'' from which we aim to select data points near the boundary. We display the \gls{mse} \gls{alc} curves, \gls{ateer} \gls{alc} curves, and the expected number of events curve at the end of all maintenance cycles in \cref{fig:example_ee,fig:example_mse}. These figures are based on a single realization of the experiment settings with parameters \( k = 1.3, \lambda = 0.1, c_1 = 0.2, c_2 = 0.6, \Delta t = 5.0, J = 50, \text{ and } B = 5.0 \) under the subset configuration.

\begin{figure}[h!]
    \centering
    \includegraphics[width=1.02\linewidth]{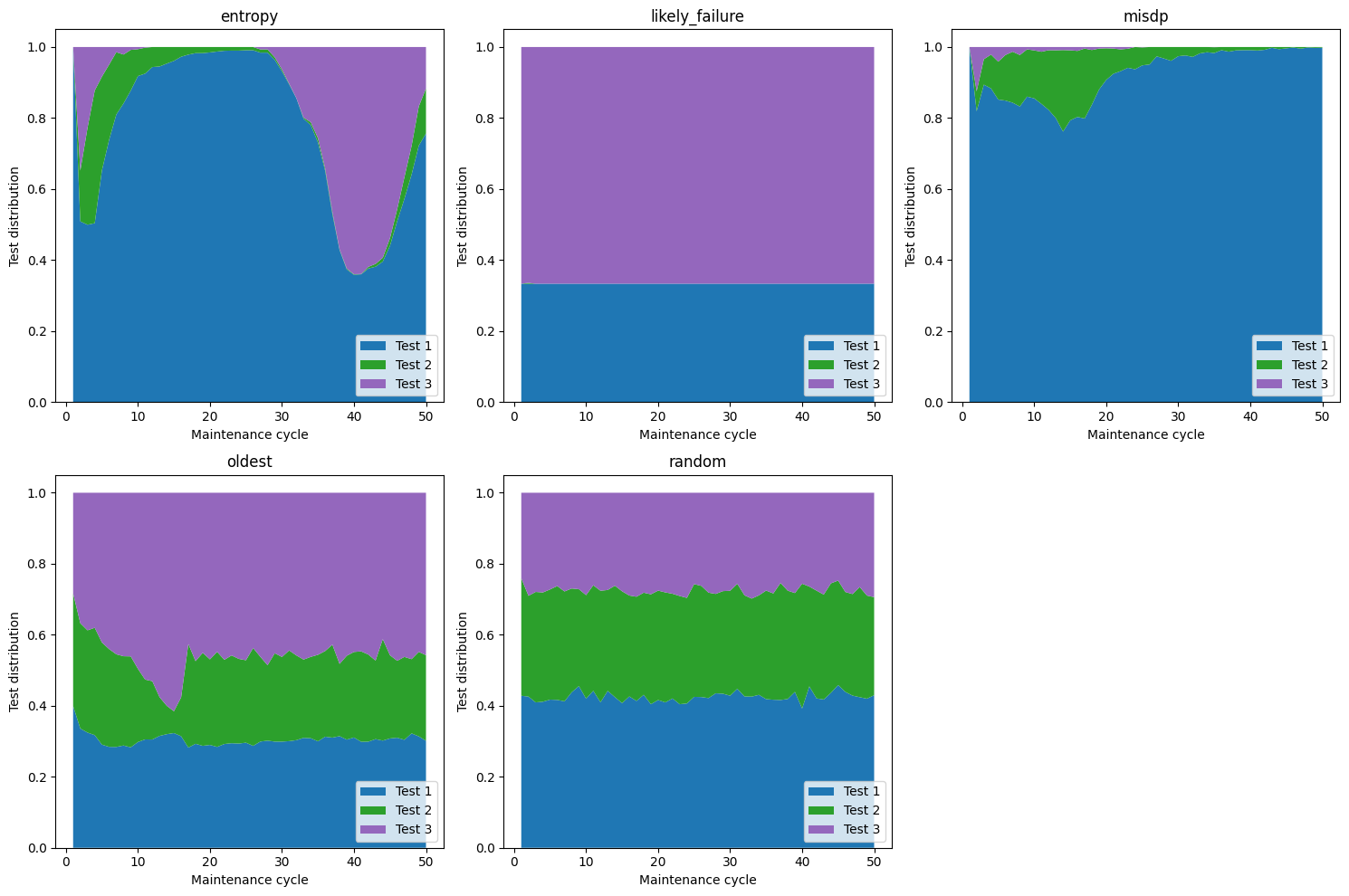}
    \caption{A stack plot illustrating the proportion of selected diagnostic tests—partial test 1, partial test 2, and proof test—at each maintenance cycle under the experiment settings \( k = 1.3 \), \( \lambda = 0.1 \), \( c_1 = 0.2 \), \( c_2 = 0.6 \), \( \Delta t = 5.0 \), \( J = 50 \), and \( B = 5.0 \) in the subset configuration. Partial test 1 = test 1, partial test 2 = test 2, and proof test = full test.}
    \label{fig:stackplots}
\end{figure}

Additionally, for the given experiment settings, we use stack plots to visualize the percentage of each diagnostic test—partial test 1, partial test 2, or proof test—performed at each maintenance cycle, illustrating the behavior of different \glspl{af} (\cref{fig:stackplots}). The Entropy \gls{af} generally appears periodic, with it alternating between using partial test 2 (test 2) or proof test (test 3) more throughout the maintenance cycles. The Most Likely Failure \gls{af} typically selects the proof test, as the reliability of a hardware system is always lower than that of any subset of its subsystems. However, due to budget constraints, not all diagnostic tests can be proof tests, resulting in some diagnostic tests being partial test 1 (test 1). The Oldest Subsystem \gls{af} closely resembles the random \gls{af} in that the proportion of selected diagnostic tests at each maintenance cycle is nearly evenly distributed among all options. However, the hardware system instances and corresponding diagnostic tests selected differ significantly by definition of the \glspl{af}.

Lastly, we note that even if the DC value is unknown, this analysis provides valuable insights into simulating possible maintenance schedules depending on hypothetical DC values. For example, given hypothetical DC values, we can answer questions such as “How many maintenance cycles or how much money should we use to achieve an MSE of X on the reliability model parameters if the DC is Y”, or “What do the system reliability plots look like under a specific maintenance plan when DC = X?” 

\glsresetall
\section{Conclusion}
\label{sec:conclusion}
We designed an improved \gls{al} \gls{af} as a relaxed \gls{misdp} to select pairs of hardware system instances and diagnostic tests, inferring reliability model parameters with minimal maintenance cycles while incorporating \gls{dc} and budget constraints.  Our method statistically ranked best amongst other \glspl{af}, such as Entropy, with respect to metrics such as   \gls{alc} \gls{ateer} and \gls{alc} \gls{mse}. We verified these results with a Friedman hypothesis test. For future work, we plan to analyze model mispecification between the data generating process and the assumed reliability model, as well as the impact of imperfect or absent \gls{dc} estimation.

\bibliographystyle{IEEEtran}  
\bibliography{bibliography.bib} 

\begin{IEEEbiography}[{\includegraphics[width=0.6in,height=1.25in,clip,keepaspectratio]{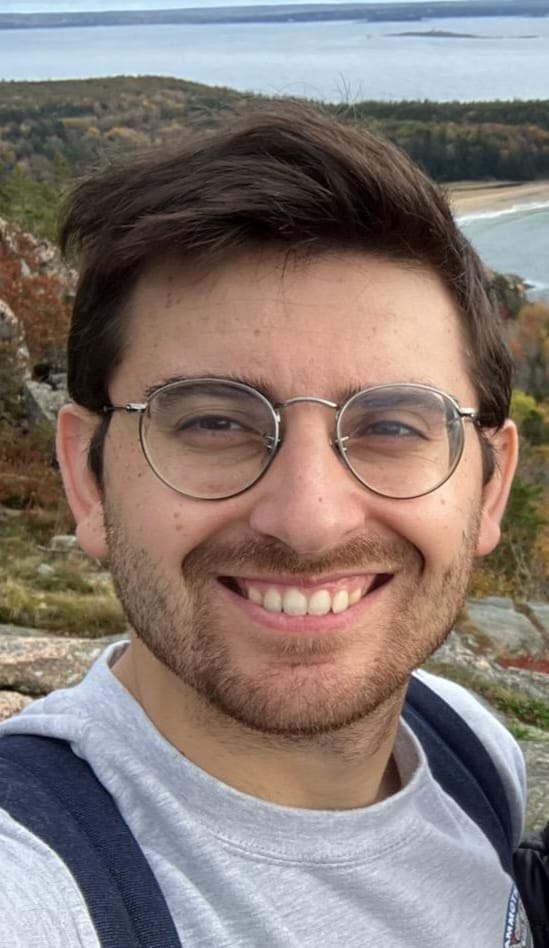}}]{Michael Potter} is a Ph.D. student at Northeastern University (NEU) advised by Deniz Erdo\u{g}mu\c{s} of the Cognitive Systems Laboratory (CSL). He received his B.S, M.S., and M.S.  degrees in Electrical and Computer Engineering from NEU and University of California Los Angeles (UCLA) in 2020, 2020, and 2022 respectively. His research interests are Bayesian Neural Networks, uncertainty quantification, and dynamics based manifold learning. 
\end{IEEEbiography}%
\vspace{-1em}

\begin{IEEEbiography}[{\includegraphics[width=1.0in,height=1.25in,clip,keepaspectratio]{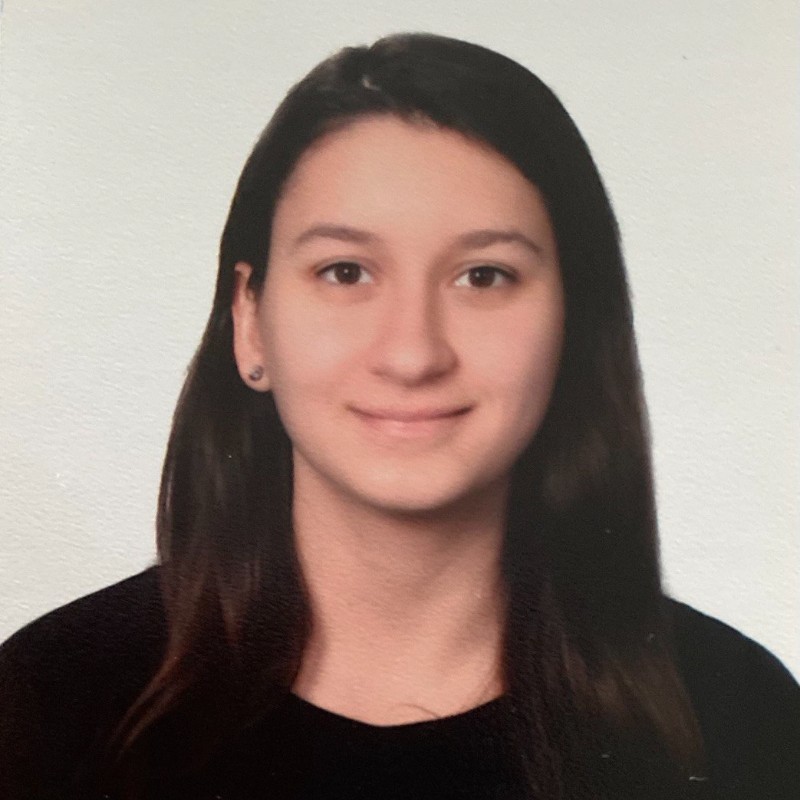}}]{Beyza Kalkanli } is a Ph.D. candidate in Computer Engineering at Northeastern University, advised by Prof. Deniz Erdogmus in the Cognitive Systems Lab (CSL). She earned her B.S. in Computer Science from Bilkent University in 2020 and her M.S. in Electrical and Computer Engineering from Northeastern University in 2022. Her research focuses on optimizing machine learning efficiency through active learning, active class selection, and domain adaptation techniques.
\end{IEEEbiography}%
\vspace{-1em}

\begin{IEEEbiography}[{\includegraphics[width=1.0in,height=1.25in,clip,keepaspectratio]{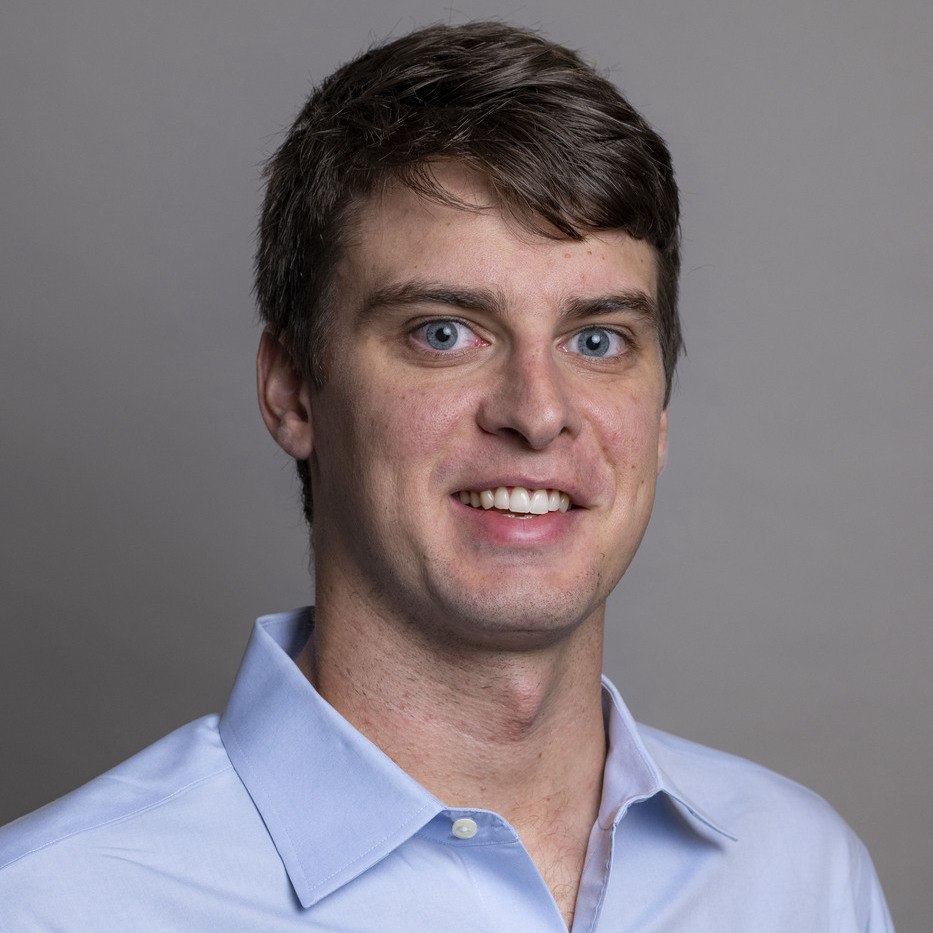}}]{Michael Everett} received the S.B., S.M., and Ph.D. degrees in mechanical engineering from the Massachusetts Institute of Technology (MIT), Cambridge, MA, USA, in 2015, 2017, and 2020, respectively. He was a Post-Doctoral Associate and Research Scientist in the Department of Aeronautics and Astronautics at MIT. He was a Visiting Faculty Researcher at Google Research.
He joined Northeastern University in 2023, where he is currently an Assistant Professor in the Department of Electrical \& Computer 
Engineering and Khoury College of Computer Sciences at Northeastern University, Boston, MA, USA. His research lies at the intersection of machine learning, robotics, and control theory, with specific interests in the theory and application of safe and robust neural feedback loops.
Dr. Everett's work has been recognized with numerous awards, including the Best Paper Award in Cognitive Robotics at IEEE/RSJ International Conference on Intelligent Robots and Systems (IROS) 2019.
\end{IEEEbiography}
\vspace{-1em}

\begin{IEEEbiography}
[{\includegraphics[width=1in,height=1.25in,clip,keepaspectratio]{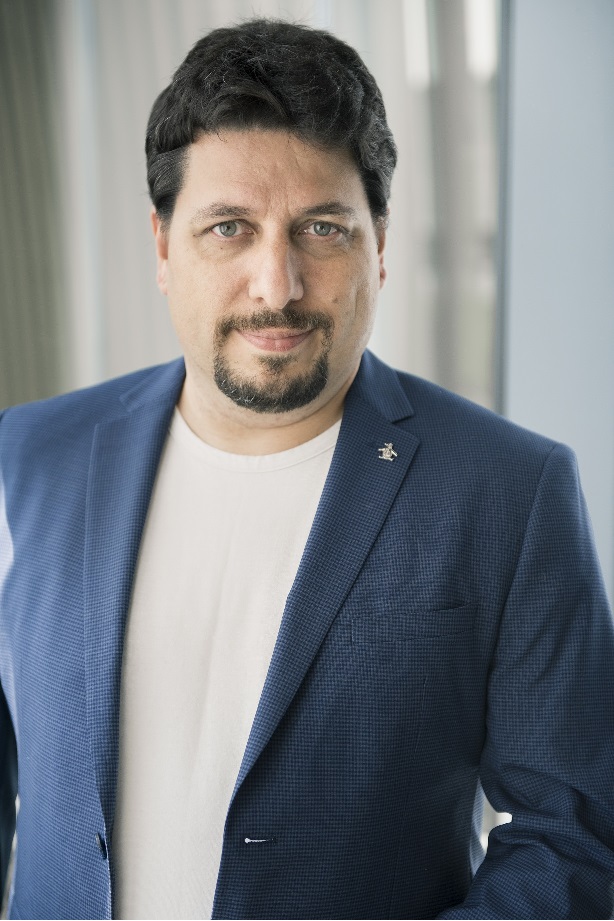}}]
{Deniz Erdo\u{g}mu\c{s}} (Sr Member, IEEE), received BS in EE and Mathematics (1997), and MS in EE (1999) from the Middle East Technical University, PhD in ECE (2002) from the University of Florida, where he was a postdoc until 2004. He was with CSEE and BME Departments at OHSU (2004-2008). Since 2008, he has been with the ECE Department at Northeastern University. His research focuses on statistical signal processing and machine learning with applications data analysis, human-cyber-physical systems, sensor fusion and intent inference for autonomy. He has served as associate editor and technical committee member for multiple IEEE societies.
\end{IEEEbiography}

\appendix

\begin{appendices}  
\onecolumn

\section{Overlap Partial Test} 
\label{appsec:reliability_derivation}

\subsection{Re-expressing subsystem hazards}
By rearranging the \glspl{dc} equations, and the fact that  the total intensity rate is the sum of the subsystem hazard rate, we have a system of 3 equations with 3 unknowns:
\begin{align}
    c_1 h(t) &= h_1(t) + h_2(t) \\
    c_2 h(t) &= h_2(t) + h_3(t) \\
    h(t) &= h_1(t) + h_2(t) + h_3(t) \label{eqn:total_failure}
\end{align}

Therefore, we may solve for $h_1(t),h_2(t)$ and $h_3(t)$ in terms of $c_1,c_2$ and $h(x,t)$.

\subsection{after partial test 1}
\begin{align}
    p(\min(T_1,T_2) > t_{age} |  T_1 > t_{agelt1},T_2 > t_{agelt}, T_3 > t_{agelt3}) & = \\
    \frac{p(\min(T_1,T_2) > t_{age}, T_1 > t_{agelt1},T_2 > t_{agelt}, T_3 > t_{agelt3})}{p(T_1 > t_{agelt1}, T_2 > t_{agelt}, T_3 > t_{agelt3})} & =\\
    \frac{p(T_1 > t_{age}, T_2>t_{age}, T_1 > t_{agelt1},T_2 > t_{agelt}, T_3 > t_{agelt3})}{p(T_1 > t_{agelt1}, T_2 > t_{agelt}, T_3 > t_{agelt3})} & =\\
    \frac{p(T_1 >  t_{age}, T_2 > t_{age}, T_3 > t_{agelt3})}{p(T_1 > t_{agelt1}, T_2 > t_{agelt}, T_3 > t_{agelt3})} & =\\
    \frac{p(T_1>t_{age})p(T_2>t_{age})p(T_3>t_{agelt3})}{p(T_1>t_{agelt1})p(T_2>t_{agelt})p(T_3>t_{agelt3})} & =\\
    \frac{p(T_1>t_{age})p(T_2>t_{age})}{p(T_1>t_{agelt1})p(T_2>t_{agelt})} & =
    \frac{e^{-\alpha_1t_{age}^{k}}e^{-\alpha_2t_{age}^{k}}}{e^{-\alpha_1t_{agelt1}^{k}}e^{-\alpha_2t_{agelt}^{k}}} \\
    &= e^{- \alpha \cdot (1-c_2)(t_{age}^{k} - t_{agelt1}^{k}) -\alpha \cdot (c_1+c_2 - 1)(t_{age}^{k} - t_{agelt}^{k})} \\
    &= e^{-\alpha \left[c_1 t_{age}^{k} - (1 - c_2) t_{agelt1}^{k} + (1-c_1 - c_2) t_{agelt}^{k}\right]}
\end{align}
\subsection{after partial test 2}
The same calculations for partial test 1  are  used for partial test 2 to give
\begin{align}
    p(\min(T_2,T_3) > t_{age} |  T_1 > t_{agelt1},T_2 > t_{agelt}, T_3 > t_{agelt3}) & = \\
    &= e^{- \alpha \cdot (1-c_1)(t_{age}^{k} - t_{agelt3}^{k}) -\alpha \cdot (c_1+c_2 - 1)(t_{age}^{k} - t_{agelt}^{k})} \\
    &= e^{-\alpha \left[c_2 t_{age}^{k} - (1 - c_1) t_{agelt3}^{k} + (1 - c_1 - c_2) t_{agelt}^{k}\right]}
\end{align}

\subsection{after partial test 1 and partial test 2}
\begin{align}
    p(\min(T_1,T_2,T_3) > t_{age} |  T_1 > t_{agelt1},T_2 > t_{agelt}, T_3 > t_{agelt3}) & = \\    
    \frac{p(T_1>t_{age})p(T_2>t_{age})p(T_3>t_{age})}{p(T_1>t_{agelt1})p(T_2>t_{agelt})p(T_3>t_{agelt3})} & =\\
    =e^{- \alpha \cdot (1-c_2)(t_{age}^{k} - t_{agelt1}^{k})}e^{-\alpha \cdot (c_1+c_2 - 1)(t_{age}^{k} - t_{agelt}^{k}) } e^{- \alpha \cdot (1-c_1)(t_{age}^{k} - t_{agelt3}^{k})} & \\
    =e^{-\alpha \left[t_{age}^{k} - (1 - c_2) t_{agelt1}^{k} + (1 - c_1 -c2) t_{agelt}^{k} - (1 - c_1) t_{agelt3}^{k}\right]} 
\end{align}

\section{Subset Partial Test} 
\label{appsec:reliability_subset_derivation}

\subsection{Re-expressing subsystem hazards}
By rearranging the \glspl{dc} equations, and the fact that  the total intensity rate is the sum of the subsystem hazard rate, we have a system of 3 equations with 3 unknowns:
\begin{align}
    c_1 h(t) &= h_1(t) \\
    c_2 h(t) &= h_1(t) + h_2(t) \\
    h(t) &= h_1(t) + h_2(t) + h_3(t)
\end{align}

Therefore, we may solve for $h_1(t),h_2(t)$ and $h_3(t)$ in terms of $c_1,c_2$ and $h(x,t)$.

\subsection{after partial test 2}
\begin{align}
    p(\min(T_1,T_2) > t_{age} |  T_1 > t_{agelt},T_2 > t_{agelt2}, T_3 > t_{agelt3}) & = \\
    \frac{p(\min(T_1,T_2) > t_{age}, T_1 > t_{agelt},T_2 > t_{agelt2}, T_3 > t_{agelt3})}{p(T_1 > t_{agelt}, T_2 > t_{agelt2}, T_3 > t_{agelt3})} & =\\
    \frac{p(T_1 > t_{age}, T_2>t_{age}, T_1 > t_{agelt},T_2 > t_{agelt2}, T_3 > t_{agelt3})}{p(T_1 > t_{agelt}, T_2 > t_{agelt2}, T_3 > t_{agelt3})} & =\\
    \frac{p(T_1 >  t_{age}, T_2 > t_{age}, T_3 > t_{agelt3})}{p(T_1 > t_{agelt}, T_2 > t_{agelt2}, T_3 > t_{agelt3})} & =\\
    \frac{p(T_1>t_{age})p(T_2>t_{age})p(T_3>t_{agelt3})}{p(T_1>t_{agelt})p(T_2>t_{agelt2})p(T_3>t_{agelt3})} & =\\
    \frac{p(T_1>t_{age})p(T_2>t_{age})}{p(T_1>t_{agelt})p(T_2>t_{agelt2})} & =
    \frac{e^{-\alpha_1t_{age}^{k}}e^{-\alpha_2t_{age}^{k}}}{e^{-\alpha_1t_{agelt}^{k}}e^{-\alpha_2t_{agelt2}^{k}}} \\
    &= e^{- \alpha \cdot c_1 (t_{age}^{k} - t_{agelt}^{k}) -\alpha \cdot (c_2-c_1)(t_{age}^{k} - t_{agelt2}^{k})} \\
\end{align}
\subsection{after partial test 1}
The same calculations for partial test 1  are  used for partial test 2 to give
\begin{align}
    p(T_1 > t_{age} |  T_1 > t_{agelt},T_2 > t_{agelt2}, T_3 > t_{agelt3}) & = \\
    \frac{p(T_1 > t_{age}, T_1 > t_{agelt},T_2 > t_{agelt2}, T_3 > t_{agelt3})}{p(T_1 > t_{agelt}, T_2 > t_{agelt2}, T_3 > t_{agelt3})} & =\\
    \frac{p(T_1 > t_{age}, T_1 > t_{agelt},T_2 > t_{agelt2}, T_3 > t_{agelt3})}{p(T_1 > t_{agelt}, T_2 > t_{agelt2}, T_3 > t_{agelt3})} & =\\
    \frac{p(T_1 >  t_{age}, T_2 > t_{agelt2}, T_3 > t_{agelt3})}{p(T_1 > t_{agelt}, T_2 > t_{agelt2}, T_3 > t_{agelt3})} & =\\
    \frac{p(T_1>t_{age})p(T_2>t_{agelt2})p(T_3>t_{agelt3})}{p(T_1>t_{agelt})p(T_2>t_{agelt2})p(T_3>t_{agelt3})} & =\\
    \frac{p(T_1>t_{age})}{p(T_1>t_{agelt})} & =
    \frac{e^{-\alpha_1t_{age}^{k}}}{e^{-\alpha_1t_{agelt}^{k}}} \\
    &= e^{- \alpha \cdot c_1 (t_{age}^{k} - t_{agelt}^{k})}
\end{align}

\subsection{after partial test 1 and partial test 2}
\begin{align}
    p(\min(T_1,T_2,T_3) > t_{age} |  T_1 > t_{agelt},T_2 > t_{agelt2}, T_3 > t_{agelt3}) & = \\    
    \frac{p(T_1>t_{age})p(T_2>t_{age})p(T_3>t_{age})}{p(T_1>t_{agelt})p(T_2>t_{agelt2})p(T_3>t_{agelt3})} & =\\
    = e^{- \alpha \cdot c_1 (t_{age}^{k} - t_{agelt}^{k}) -\alpha \cdot (c_2-c_1)(t_{age}^{k} - t_{agelt2}^{k})-\alpha (1-c_2) (t_{age}^k - t_{agelt3}^k)} & \\
\end{align}

\section{Conditional Fisher Information} 
\subsection{Overlap partial test}
\tiny
\begin{align}
    R(t|PT_1,PT_2,PT_3) &= e^{- \alpha  \left[PT_1 (1-c_2) (t_{age}^{k} - t_{agelt1}^{k}) + PT_2 (c_2+c_1-1)(t_{age}^{k} - t_{agelt}^{k}) + PT_3 (1-c_1) (t_{age}^k - t_{agelt3}^k)\right]} \\
    LL &= (1-R(t|PT_1,PT_2))^{y} R(t|PT_1,PT_2)^{(1-y)} \\
    \log(LL) &= y \log\left(1-e^{- \alpha  \left[PT_1 (1-c_2) (t_{age}^{k} - t_{agelt1}^{k}) + PT_2 (c_2+c_1-1)(t_{age}^{k} - t_{agelt}^{k}) + PT_3 (1-c_1) (t_{age}^k - t_{agelt3}^k)\right]}\right)  \\
    & \qquad - (1-y)\times\alpha  \left[PT_1 (1-c_2) (t_{age}^{k} - t_{agelt1}^{k}) + PT_2 (c_2+c_1-1)(t_{age}^{k} - t_{agelt}^{k}) + PT_3 (1-c_1) (t_{age}^k - t_{agelt3}^k)\right]
\end{align}

\begin{align}
    \frac{\partial \log(LL)}{\partial \alpha} &= -y \frac{R(t\mid \ldots)}{1-R(t \mid \ldots)}   \left[PT_1 (1-c_2) (t_{age}^{k} - t_{agelt1}^{k}) + PT_2 (c_2+c_1-1)(t_{age}^{k} - t_{agelt}^{k}) + PT_3 (1-c_1) (t_{age}^k - t_{agelt3}^k)\right]  \\
    & \qquad - (1-y)\times  \left[PT_1 (1-c_2) (t_{age}^{k} - t_{agelt1}^{k}) + PT_2 (c_2+c_1-1)(t_{age}^{k} - t_{agelt}^{k}) + PT_3 (1-c_1) (t_{age}^k - t_{agelt3}^k)\right] \\
    & =\left[-y \frac{R(t\mid \ldots)}{1-R(t \mid \ldots)} - (1-y) \right]  \left[PT_1 (1-c_2) (t_{age}^{k} - t_{agelt1}^{k}) + PT_2 (c_2+c_1-1)(t_{age}^{k} - t_{agelt}^{k}) + PT_3 (1-c_1) (t_{age}^k - t_{agelt3}^k)\right] \\
    \frac{\partial \log(LL)}{\partial k} &= \left[-y \frac{R(t\mid \ldots)}{1-R(t \mid \ldots)} - (1-y) \right] \alpha  \biggr[PT_1 (1-c_2) \left(t_{age}^{k} \log(t_{age}) - t_{agelt1}^{k} \log(t_{agelt1})\right) \\
    & \qquad \qquad \qquad \qquad \qquad \qquad  \qquad \qquad \qquad + PT_2 (c_2+c_1-1)\left(t_{age}^{k} \log(t_{age}) - t_{agelt}^{k} \log(t_{agelt}) \right) \\
    & \qquad \qquad \qquad \qquad \qquad \qquad  \qquad \qquad \qquad + PT_3 (1-c_1) \left(t_{age}^k \log(t_{age}) - t_{agelt3}^k \log(t_{agelt3})\right)\biggr] \\
\end{align}

\subsection{Subset partial test}
\tiny
\begin{align}
    R(t|PT_1,PT_2,PT_3) &= e^{- \alpha  \left[PT_1  c_1 (t_{age}^{k} - t_{agelt}^{k}) +  PT_2 (c_2-c_1)(t_{age}^{k} - t_{agelt2}^{k}) + PT_3 (1-c_2) (t_{age}^k - t_{agelt3}^k)\right]} 
\end{align}
similarly to above, we may express the partial derivatives of the log likelihood as
\begin{align}
    \frac{\partial \log(LL)}{\partial \alpha} &= -y \frac{R(t\mid \ldots)}{1-R(t \mid \ldots)}   \left[PT_1 c_1 (t_{age}^{k} - t_{agelt}^{k}) + PT_2 (c_2-c_1)(t_{age}^{k} - t_{agelt2}^{k}) + PT_3 (1-c_2) (t_{age}^k - t_{agelt3}^k)\right]  \\
    & \qquad - (1-y)\times  \left[PT_1 c_1 (t_{age}^{k} - t_{agelt}^{k}) + PT_2 (c_2-c_1)(t_{age}^{k} - t_{agelt2}^{k}) + PT_3 (1-c_2) (t_{age}^k - t_{agelt3}^k)\right] \\
    & =\left[-y \frac{R(t\mid \ldots)}{1-R(t \mid \ldots)} - (1-y) \right]  \left[PT_1 c_1 (t_{age}^{k} - t_{agelt}^{k}) + PT_2 (c_2-c_1)(t_{age}^{k} - t_{agelt2}^{k}) + PT_3 (1-c_2) (t_{age}^k - t_{agelt3}^k)\right] \\
    \frac{\partial \log(LL)}{\partial k} &= \left[-y \frac{R(t\mid \ldots)}{1-R(t \mid \ldots)} - (1-y) \right] \alpha  \biggr[PT_1 c_1 \left(t_{age}^{k} \log(t_{age}) - t_{agelt}^{k} \log(t_{agelt})\right) \\
    & \qquad \qquad \qquad \qquad \qquad \qquad  \qquad \qquad \qquad + PT_2 (c_2 - c_1)\left(t_{age}^{k} \log(t_{age}) - t_{agelt2}^{k} \log(t_{agelt2}) \right) \\
    & \qquad \qquad \qquad \qquad \qquad \qquad  \qquad \qquad \qquad + PT_2 (1-c_2) \left(t_{age}^k \log(t_{age}) - t_{agelt3}^k \log(t_{agelt3})\right)\biggr] \\
\end{align}

\end{appendices}

\EOD

\end{document}